\documentclass[journal]{IEEEtran}
\newif\ifdouble 
\newif\ifsingle 
 	\doubletrue
 	\singlefalse 
\usepackage[pdftex]{graphicx}
\usepackage{epstopdf}
\usepackage[cmex10]{amsmath}
\usepackage{ifthen}
\usepackage{commath}
\usepackage{amsfonts, amssymb}
\usepackage{esint}
\usepackage{enumerate}
\usepackage{cite} 
\usepackage{hyperref}
\usepackage{multirow}
\usepackage{color}
\usepackage{balance}
\makeatletter
\long\def\@makecaption#1#2{\ifx\@captype\@IEEEtablestring%
\footnotesize\begin{center}{\normalfont\footnotesize #1}\\
{\normalfont\footnotesize\scshape #2}\end{center}%
\@IEEEtablecaptionsepspace
\else
\@IEEEfigurecaptionsepspace
\setbox\@tempboxa\hbox{\normalfont\footnotesize {#1.}~~ #2}%
\ifdim \wd\@tempboxa >\hsize%
\setbox\@tempboxa\hbox{\normalfont\footnotesize {#1.}~~ }%
\parbox[t]{\hsize}{\normalfont\footnotesize \noindent\unhbox\@tempboxa#2}%
\else
\hbox to\hsize{\normalfont\footnotesize\hfil\box\@tempboxa\hfil}\fi\fi}
\makeatother

\newtheorem{theorem}{Theorem}
\newtheorem{corollary}{Corollary}

\newcommand{\kf}{k_f} 
\newcommand{\kfm}{k_{f^{\star}}} 
\newcommand{\kb}{k_b}
\newcommand{\kd}{k_d}
\newcommand{\pr}{\text{Pr}}
\newcommand{\w}{\text{W}}
\newcommand{\erf}{\mathrm{erf}\,}
\newcommand{\erfc}{\mathrm{erfc}\,} 
\newcommand{\pacv}[1]{P_{AC}({#1}|\vec{r}_0)}
\newcommand{\pac}[1]{P_{AC}({#1}|r_0)}
\newcommand{\pav}[2]{P_{A}({#1},{#2}|\vec{r}_0)}
\newcommand{\pa}[2]{P_{A}({#1},{#2}|r_0)} 
\newcommand{\hs}{\hspace*{-0.75 mm}} 
\newcommand{\di}[1]{{#1}^{\prime}}
\newcommand{\padi}[2]{P_{A}^{\prime}(\di{#1},\di{#2}|\di{r}_0)} 
\newcommand{\pacdi}[1]{P_{AC}^{\prime}(\di{#1}|\di{r}_0)} 
\newcommand{\dkf}{\di{k}_f} 
\newcommand{\dkb}{\di{k}_b}
\newcommand{\dkd}{\di{k}_d} 
\newcommand{\dkfm}{\di{k}_{f^{\star}}} 
\newcommand{\jsmr}{J_{Sph}^{M}} 
\newcommand{\js}{J_{Sph}}
\newcommand{\jr}{J_{Rec}} 
\newcommand{\pacm}[1]{P_{AC}^{M}({#1}|r_0)}
\newcommand{\pacn}[1]{P_{AC}^{\star}({#1}|r_0)}
\allowdisplaybreaks
\begin{document}

\title{Comprehensive Reactive Receiver Modeling for Diffusive Molecular Communication Systems: Reversible Binding, Molecule Degradation, and Finite Number of Receptors}
\author{Arman~Ahmadzadeh, ~\IEEEmembership{Student Member,~IEEE,}
        Hamidreza Arjmandi, Andreas Burkovski, 
        and~Robert~Schober,~\IEEEmembership{Fellow,~IEEE}
\thanks{Manuscript received June 03, 2016; accepted August 23, 2016. This paper was presented in part at the International Conference on Communications (ICC) 2016, Kuala Lumpur, Malaysia \cite{Arman3}. The associate editor coordinating the review of this manuscript and approving it for publication was Prof. Henry Hess.}        
\thanks{A. Ahmadzadeh and R. Schober are with the Institute for Digital Communication, University of Erlangen-Nuremberg, Erlangen, Germany (email: \{arman.ahmadzadeh, robert.schober\}@fau.de).}
\thanks{A. Burkovski is with the Institute for Microbiology, University of Erlangen-Nuremberg, Erlangen, Germany (email: andreas.burkovski@fau.de).}
\thanks{H. Arjmandi is with the Department of Electrical Engineering, Sharif University of Technology, Tehran, Iran. (email: arjmandi@ee.sharif.edu).}} 

\maketitle 
\begin{abstract} 
This paper studies the problem of receiver modeling in molecular communication systems. We consider the diffusive molecular communication channel between a transmitter nano-machine and a receiver nano-machine in a fluid environment. The information molecules released by the transmitter nano-machine into the environment can degrade in the channel via a first-order degradation reaction and those that reach the receiver nano-machine can participate in a reversible bimolecular reaction with receiver receptor proteins. Thereby, we distinguish between two scenarios. In the first scenario, we assume that the entire surface of the receiver is covered by receptor molecules. We derive a closed-form analytical expression for the expected received signal at the receiver, i.e., the expected number of activated receptors on the surface of the receiver. Then, in the second scenario, we consider the case where the number of receptor molecules is finite and the uniformly distributed receptor molecules cover the receiver surface only partially. We show that the expected received signal for this scenario can be accurately approximated by the expected received signal for the first scenario after appropriately modifying the forward reaction rate constant. The accuracy of the derived analytical results is verified by Brownian motion particle-based simulations of the considered environment, where we also show the impact of the effect of receptor occupancy on the derived analytical results.          
\end{abstract}

\section{Introduction} 
In nature, one of the primary means of communication among biological entities, ranging from organelles to organisms, is molecular communication (MC), where molecules are the carriers of information. MC is also an attractive option for the design of intelligent synthetic communication systems at nano- and micro-scale. Sophisticated synthetic MC systems are expected to have various biomedical, environmental, and industrial applications \cite{NakanoB}.    

Similar to any other communication system, the design of basic functionalities such as modulation, detection, and estimation requires accurate models for the transmitter, channel, and receiver. However, the modeling of these components in MC systems is vastly different from the modeling of traditional communication systems as the size of the nodes of MC systems is on the order of tens of nanometer to tens of micrometer \cite{NakanoB}. At nano- and micro-scale, materials show different physical, chemical, electrical, and magnetic behaviour which has to be carefully accounted for in the modeling, fabrication, and development of new devices, materials, and systems. 
Furthermore, in MC systems, the communication environment, i.e., the channel, and the pertinent impairments are completely different from those in traditional communication systems. For example, in a typical MC setup, the transmitter and the receiver are suspended in a \emph{fluid} environment. Moreover, in diffusion-based MC, the transportation of the information-carrying signalling molecules relies on free diffusion; no additional infrastructure is required. Additionally, the signalling molecules may be affected by various environmental effects such as \emph{flow} and/or \emph{chemical reactions}, which may prevent the molecules from reaching the receiver \cite{NakanoB}.

The molecules that succeed in reaching the receiver may react with the receptors on the surface of the receiver and activate them. This is a common reception mechanism in natural biological cells \cite{AlbertsBook}. The number of activated receptors can be interpreted as the received signal. The received signal is inevitably corrupted by noise that arises from the stochastic arrival of the molecules by diffusion and from the stochastic binding of the molecules to the receptors. Once an information-carrying molecule binds to a receptor, i.e., \emph{activates the receptor}, this receptor transduces the received noisy message into a cell response via a set of signaling pathways, see \cite[Chapter 16]{AlbertsBook}. Thus, having a meaningful model for the reception process that captures the effects of the main phenomena occurring in the channel and at the receiver is of particular importance for the design of synthetic MC systems and is the focus of this paper.

The modeling of the \emph{reception mechanism} at the receiver has been studied extensively in the existing MC literature; see \cite{PierobonJ1, PierobonJ2, MahfuzJ1, PierobonJ3, ShahMohammadianProc1, YilmazL1, Akkaya, Heren, ChunJ1, Yansha1}. Most of these works assume that the receiver is transparent and the received signal is approximated by the local concentration of the information-carrying molecules inside the receiver, see \cite{PierobonJ1, PierobonJ2, MahfuzJ1}. However, this approach neglects the impact of the chemical reactions at the receiver surface required for sensing the concentration of molecules in its vicinity. In a first attempt to take the effect of chemical reactions at the receiver into account, the authors in \cite{PierobonJ3} and \cite{ShahMohammadianProc1} modelled the reception mechanism at the receiver using the theory of ligand-receptor binding kinetics, where molecules (ligands) released by a transmitter can reversibly react with receptors at the receiver surface and produce output molecules whose concentration is then referred to as the output signal in \cite{PierobonJ3}, \cite{ShahMohammadianProc1}. Analytical expressions that relate the output signal to the concentration of the molecules at the receiver were derived. However, in \cite{PierobonJ3} and \cite{ShahMohammadianProc1}, the authors assumed the diffusion of molecules to be independent from the reaction mechanisms at the receiver, i.e., the equations describing the output signal were derived for a given concentration at the receiver. This assumption may not be justified as the ligand concentration close to the receiver is subject to fluctuation due to the association and dissociation processes at the receiver surface. Hence, the reaction-diffusion equation is highly \emph{coupled} and cannot be separated. The authors in \cite{YilmazL1} approximated the reception mechanism by modeling the receiver as a fully-absorbing sphere, where molecules released by the transmitter are absorbed by the receiver as soon as they hit its surface, and derived an analytical time domain expression for the number of absorbed molecules. In particular, this assumption is equivalent to a second-order \emph{irreversible} reaction mechanism on the surface of the receiver where the reaction rate constant is \emph{infinite}. This assumption implies that the effect of the \emph{reaction mechanism}, upon the contact of a released molecule with the surface of the receiver, is deterministic. However, it may not be realistic to assume that every collision of a molecule with the receiver surface triggers a reaction, since the reaction mechanism is a stochastic process. This model was extended in \cite{Akkaya} and \cite{Heren} to incorporate the effects of a finite number of receptors and a first-order degradation reaction in the channel, respectively. The author in \cite{ChunJ1} approximated the reception mechanism by a \emph{first-order reversible} reaction inside the volume of the receiver, and a reaction-diffusion master equation (RDME) model was used to solve the corresponding coupled reaction-diffusion equation. An analytical expression for the expected number of output molecules was derived in the Laplace domain. However, a time-domain solution was not provided. In the recent work \cite{Yansha1}, the authors modelled the reception mechanism as a \emph{second-order reversible} reaction on the surface of the receiver assuming that the whole surface of the receiver is reactive, and derived an analytical time domain expression for the number of absorbed molecules. However, this expression required numerical integration.     

In this paper, we model the reception mechanism at the receiver surface as a second-order reversible reaction, where an information molecule released by the transmitter can \emph{reversibly} react with \emph{receptor} protein molecules covering the surface of the receiver and activate them. The number of activated receptors can be interpreted as the received signal and later be used for detection. Unlike \cite{PierobonJ1, PierobonJ2, MahfuzJ1, PierobonJ3, ShahMohammadianProc1, YilmazL1, Akkaya, ChunJ1, Yansha1}, we take into account that in a realistic environment, molecules released by the transmitter may degrade in the channel via a first-order degradation reaction. The concentration of information molecules close to the receiver and, as a result, the number of activated receptors are subject to fluctuations caused by the degradation of information molecules in the channel and the association and dissociation of information molecules with receptor protein molecules. In order to accurately account for both reaction mechanisms in our model, we solve the coupled reaction-diffusion equation capturing the diffusion and degradation of information-carrying molecules in the channel \emph{jointly} with boundary conditions capturing the effect of the reversible reaction of the information molecules on the surface of the receiver with receptor protein molecules. Unlike \cite{ChunJ1, Yansha1}, we derive a \emph{closed-form time domain} expression for the expected number of activated receptor protein molecules in response to the impulsive release of molecules at the transmitter. This analytical expression constitutes a general framework for modeling the expected received signal at the receiver nano-machine. We show that the results for the expected received signal in \cite{YilmazL1, Akkaya, Heren} are special cases of our analysis. The main contributions of this paper can be summarized as follows:
\begin{enumerate}
	\item Extending our preliminary work in \cite{Arman3}, we provide a closed-form analytical expression for the expected received signal, i.e., the expected number of activated receptors, under the assumptions that the \emph{entire} surface of the receiver is covered with receptor protein molecules and that the formations of the individual ligand-receptor complexes (activated receptors) are independent of each other, i.e., multiple information molecules can react on the surface of the receiver at the same time and at the same location. These two assumptions make our analysis analytically tractable. Furthermore, the obtained expression provides the basis for the analysis of the more realistic case where the receptors cover the surface of the receiver only partially. 
	\item We provide a dimensional analysis and present the equations describing the expected received signal in a dimensionless form. The dimensional analysis generalizes our system model and reduces the number of variables that appear in the equations. 
	\item We consider the effect of individual receptor modeling and show that the expected received signal of a receiver partially covered with a \emph{finite} number of receptors can be accurately approximated by the expected received signal of a receiver that is fully covered with receptors after appropriately modifying the forward reaction rate constant. In particular, we employ a technique called boundary homogenization \cite{Berg1, Zwanzig1, Zwanzig2, Berezhkovskii1, Berezhkovskii2} and derive a closed-form expression for the effective forward reaction rate constant as a function of the system parameters. 
	\item We introduce a Brownian motion particle-based simulation framework to asses the accuracy of the derived analytical expressions for the expected received signal. Employing this simulation framework, we show the impact of receptor occupancy on the expected received signal, which was not included in our analysis and refers to the fact that each receptor can bind only to one signalling molecule at a time. The simulation framework also facilitates the reproduction of our results by other researchers in the field of MC.             
\end{enumerate}      

The rest of this paper is organized as follows. In Section \ref{Sec.SysMod}, we introduce the system model. In Section \ref{Sec.CIR}, we derive a closed-form analytical expression for the expected number of activated receptor protein molecules, which we refer to as the received signal, when receptor protein molecules cover the entire surface of the receiver. Then, in Section \ref{Sec. FiniteRecNum}, we extend the derived analytical expression for the expected received signal to the case when the number of receptors is finite. In Section \ref{Sec. SimulationFramework}, we describe the simulation framework used for the adopted particle-based simulator. In Section \ref{Sec. Simulations}, we present simulation and analytical results, and, finally, some conclusions are drawn in Section \ref{Sec.Con}.   
\section{system model} 
\label{Sec.SysMod} 
In this section, we introduce the system model considered in this paper.
\ifdouble
\begin{figure}[!t] 
	\centering
	\includegraphics[scale = 1.9]{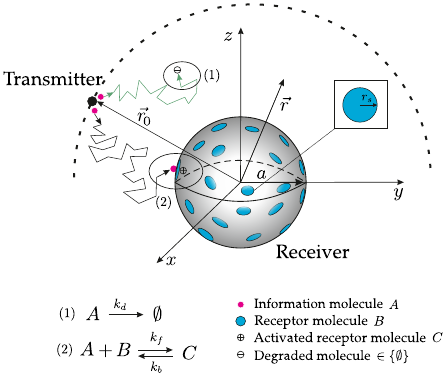}
	\caption{Schematic diagram of the considered system model, where receiver and transmitter are shown as a gray sphere in the center of the coordinate system and a black dot, respectively. The surface of the receiver is partially covered with type $B$ receptor molecules, shown in blue color, where each receptor is modeled as a circular patch with radius $r_s$. Two sample trajectories of an $A$ information molecule (denoted by a red dot) that result in a degraded molecule (circle with a ``-'' inside) and an activated receptor molecule $C$ (circle with a ``+'' inside ) are shown as  green and black arrows, respectively.} 
	\label{Fig.SystemModel}
\end{figure}
\else 
\begin{figure}[!t] 
	\centering
	\includegraphics[scale = 2.5]{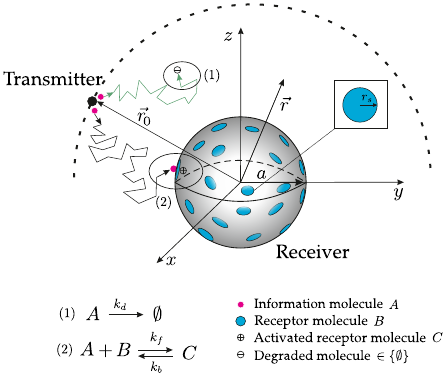}\vspace*{-2 mm}
	\caption{Schematic diagram of the considered system model, where receiver and transmitter are shown as a gray sphere in the center of the coordinate system and a black dot, respectively. The surface of the receiver is partially covered with type $B$ receptor molecules, shown in blue color, where each receptor is modeled as a circular patch with radius $r_s$. Two sample trajectories of an $A$ information molecule (denoted by a red dot) that result in a degraded molecule (circle with a ``-'' inside) and an activated receptor molecule $C$ (circle with a ``+'' inside ) are shown as  green and black arrows, respectively.}\vspace*{-5 mm} 
	\label{Fig.SystemModel}
\end{figure}
\fi
We consider an unbounded three-dimensional fluid environment with constant temperature and viscosity. We assume that the transmitter is a point source placed at position $\vec{r}_0 = (x_0,\ y_0,\ z_0)$ and the receiver is spherical in shape with a fixed radius $a$ and placed at the center of the coordinate system, see Fig. \ref{Fig.SystemModel}. 

We assume that the transmitter uses one specific type of molecule, denoted by $A$, for sending information to the receiver. Hence, we refer to the information molecules as $A$ molecules. Furthermore, we assume that the transmitter instantaneously releases a fixed number of $A$ molecules, $N_A$, at time $t_0 = 0$ into the environment. 

After their release, the information molecules diffuse in the environment into all directions with constant diffusion coefficient $D_A$. Thereby, we assume that the diffusion processes of different information molecules are independent of each other. We furthermore assume that the $A$ molecules can be degraded throughout the environment via a reaction mechanism of the form 
\begin{equation}
	\label{Eq.DegReaction} 
	A \xrightarrow{\kd} \emptyset,  
\end{equation}
where $\kd$ is the degradation reaction constant in $\text{s}^{-1}$ and $\emptyset$ is a species of molecules which is not recognized by the receiver. Eq. (\ref{Eq.DegReaction}) models a first-order reaction but can also be used to approximate higher order reactions, see e.g. \cite{Steven_Andrews, NoelJ1, ChouJ1}. 

Some of the $A$ molecules released by the transmitter may reach the receiver surface and \emph{reversibly} react with a $B$ molecule to form an activated receptor (also referred to as ligand-receptor complex molecule), denoted by $C$, via a reaction mechanism of the form 
\begin{equation}
	\label{Eq.RevReaction} 
	A + B \mathrel{\mathop{\rightleftarrows}^{\kf}_{\kb}} C,  
\end{equation}
where $\kf$ and $\kb$ are the microscopic forward reaction constant in $\text{molecule}^{-1} \text{m}^3 \text{s}^{-1}$ and the microscopic backward reaction constant in $\text{s}^{-1}$, respectively. In nature, different cell-surface receptors produce different responses inside the cell via different signaling pathways once they are activated \cite[Chapter 16]{AlbertsBook}. However, in this work, we do not focus on a specific signaling pathway. Instead, we consider the formation of the $C$ ligand-receptor complex molecules as the \emph{received signal} at the receiver, which could be used for detection of information sent by the transmitter. 

In this paper, we consider two cases regarding the coverage of the receiver surface by receptor $B$ molecules. In the first case, considered in Section \ref{Sec.CIR}, in order to make the analysis of the expected received signal at the receiver analytically tractable, we make the following simplifying assumptions:
\begin{enumerate}[(a)]
	\item We do not model individual receptors but assume that the \emph{entire} surface of the receiver is fully covered with infinitely many receptor $B$ molecules. In other words, we neglect the physical properties, such as shape, of the receptor $B$ molecules and model each receptor as a point on the surface of the receiver.  
	\item We neglect the effect of receptor occupancy, which means that multiple $A$ molecules can react at the same position on the surface of the receiver, i.e., with the same receptor $B$ molecule, to form multiple $C$ molecules. With this assumption, the formations of different $C$ molecules on the surface of the receiver become independent of each other.  
\end{enumerate}
Then, in the second case, considered in Section \ref{Sec. FiniteRecNum}, we relax assumption (a) and model each receptor individually. In particular, we assume that the receiver surface is \emph{partially} covered with $M$ uniformly distributed receptor $B$ molecules, where we model each receptor $B$ molecule as a circular patch with radius $r_s$. However, in order to keep our analysis mathematically tractable, we still neglect the effect of receptor occupancy. We show the impact of assumption (b) via particle-based simulation of the considered system in Section \ref{Sec. Simulations}.         
\section{expected received signal with full receptor coverage}
\label{Sec.CIR}  
In this section, we derive a closed-form expression for the expected received signal of a receiver whose surface is fully covered by receptors. To this end, we first formulate the problem that has to be solved to find the expected received signal. Subsequently, we derive a closed-form expression for the probability of finding an $A$ molecule, which can undergo reactions (\ref{Eq.DegReaction}) and (\ref{Eq.RevReaction}), at the position defined by vector $\vec{r}$ at time $t$, given that it was released at position $\vec{r}_0$ at time $t_0$. Finally, using this result, we derive a closed-form expression for the channel impulse response.

\subsection{Problem Formulation}
We define the \emph{expected} received signal at the receiver at time $t$, $\overline{N}_C(t)$, as the time-varying number of $C$ molecules expected on the surface of the receiver at time $t$ when the transmitter released at time $t_0$ an impulse of $N_A$ $A$ molecules into the environment. Due to the random walks of the molecules, this signal is a function of time. In this subsection, we formulate the problem that has to be solved to find $\overline{N}_C(t)$. Because of the assumption of independent movement of individual molecules, we have $\overline{N}_C(t) = N_A \pacv{t}$, where $\pacv{t}$ is the probability that a given $A$ molecule, released at $\vec{r}_0$ and time $t_0=0$, causes the formation of an activated receptor molecule $C$ on the surface of the receiver at time $t$. We refer to this probability also as the \emph{channel impulse response} of the system. 

Furthermore, the probability that a given $A$ molecule released at $\vec{r}_0$ at time $t_0=0$ is at position $\vec{r}$ at time $t$, given that this molecule may undergo either one of the two reactions introduced in (\ref{Eq.DegReaction}) and (\ref{Eq.RevReaction}) during time $t$, is denoted by $\pav{\vec{r}}{t}$. Assuming for the moment $\pav{\vec{r}}{t}$ is known, we can evaluate the incoming probability flux\footnote{We note that the probability flux refers to the flux of the position probability of a \emph{single} $A$ molecule, whereas the conventional diffusive molecule flux (used in Fick's first law of diffusion) refers to the flux of the average number of $A$ molecules. For further details, we refer the interested reader to \cite[Chapter 3]{GillespieBook}.}, $-J(\vec{r},t | \vec{r}_0)$, at the surface of the receiver by applying Einstein's theory of diffusion as \cite[Eq. (3.34)]{GillespieBook} 
\begin{equation}
	\label{Eq. ProbFlux} 
	-J(\vec{r},t | \vec{r}_0)\big|_{\vec{r} \in \Omega} = D_A \nabla \pav{\vec{r}}{t}\big|_{\vec{r} \in \Omega},
\end{equation}
where $\nabla$ is the gradient operator in spherical coordinates and $\Omega$ is the surface of the receiver. Now, given $-J(\vec{r},t | \vec{r}_0)$, $-J(\vec{r},t | \vec{r}_0) \dif  \Omega \dif t$ is the probability that a given $A$ molecule reacts with the infinitesimally small surface element $\dif \Omega$ of the receiver during infinitesimally small time $\dif t$. Integrating this function over time and the surface of the receiver yields a relationship between $\pacv{t}$ and $\pav{\vec{r}}{t}$, which can be written as \cite[Eq. (3.35)]{GillespieBook} 
\begin{equation}
	\label{Eq. Relationship} 
	\pacv{t} = - \int_{0}^{t} \oiint_{\Omega} J(\vec{r},\tau | \vec{r}_0)\cdot \dif \Omega \dif \tau .   
\end{equation} 
Thus, for evaluation of $\pacv{t}$, we first need to find $\pav{\vec{r}}{t}$.  

Clearly, the evaluation of (\ref{Eq. Relationship}), and subsequently (\ref{Eq. ProbFlux}), requires knowledge of $J(\vec{r},t | \vec{r}_0)$ only on the surface of the receiver, i.e., on $\Omega$. However, since we assume that the surface of the receiver is uniformly covered by type $B$ molecules, $J(\vec{r},t | \vec{r}_0)$ and $\pav{\vec{r}}{t}$ are only functions of the magnitude of $\vec{r}$, denoted by $r$, and not of $\vec{r}$ itself. This can also be intuitively understood from Fig. \ref{Fig.SystemModel}. To this end, let us assume that the point-source transmitter is mounted on top of a virtual sphere with radius $r_0$, where $r_0$ is the magnitude of $\vec{r}_0$. Because of symmetry, we expect that releasing a given $A$ molecule from any arbitrary point on the surface of the virtual sphere leads to the same probability of reaction on the surface of the receiver, i.e., the same $\pacv{t}$ results. However, this is only possible if $J(\vec{r},t | \vec{r}_0)$ in (\ref{Eq. Relationship}) is independent of azimuthal angle $\theta$ and polar angle $\phi$ and only depends on $r$. In the remainder of this paper, we substitute $\vec{r}$ with its magnitude $r$ in (\ref{Eq. ProbFlux}) and (\ref{Eq. Relationship}) without loss of generality. As a result, (\ref{Eq. ProbFlux}) and (\ref{Eq. Relationship}) can be combined as 
\begin{equation}
	\label{Eq. RelationshipReduced} 
	\pac{t} = \int_{0}^{t} 4\pi a^2 D_A \frac{\partial \pa{r}{\tau}}{\partial r} \bigg |_{r = a} \dif \tau.    
\end{equation}  

In the following, we are interested in formulating the problem of finding $\pa{r}{t}$ for the system model specified in Section \ref{Sec.SysMod} for a receiver with full receptor coverage. In order to do so, we start with the general form of the reaction-diffusion equation for the degradation reaction in (\ref{Eq.DegReaction}), which can be written as \cite{GrinrodB1} 
\begin{equation}
	\label{Eq. Reaction-Diffusion} 
	\frac{\partial \pa{r}{t}}{\partial t} = D_A \nabla^{2} \pa{r}{t} - k_d \pa{r}{t}, 
\end{equation}
where $\nabla^{2}$ is the Laplace operator in spherical coordinates. As discussed above, releasing a given $A$ molecule from any point on the surface of a sphere with radius $r_0$ including the point defined by $\vec{r}_0$, i.e., the actual position of the transmitter, results in the same channel impulse response, $\pac{t}$. Thus, a point source defined by $\vec{r}_0$ and a source uniformly distributed on the sphere with radius $r_0$ are equivalent in this context. As a result, releasing a given $A$ molecule at $\vec{r}_0$ at time $t_0 = 0$ can be modelled with the following initial condition 
\begin{equation}
	\label{Eq. InitialCondition} 
	\pa{r}{t \to 0} = \frac{1}{4\pi r_{0}^{2}} \delta(r - r_0),
\end{equation}
where constant $1/(4\pi r_{0}^{2})$ is a normalization factor and $\delta(\cdot)$ is the Dirac delta function. The boundary conditions of the system model for the assumed unbounded environment and the reaction mechanism in (\ref{Eq.RevReaction}) on the surface of the receiver can be written as \cite[Eqs. (3), (4)]{H.KimJ1} 
\begin{equation}
	\label{Eq. BounCon1}
	\lim_{r \to \infty} \pa{r}{t} = 0,
\end{equation} 
and 
\ifdouble 
\begin{IEEEeqnarray}{rCl}
	\label{Eq. BounCon2} 
	4\pi a^2 D_A \frac{\partial \pa{r}{t}}{\partial r} \bigg |_{r = a} & \hspace*{-1 mm} = & k_f \pa{a}{t} 	 
	 - k_b  [1 - S(t|r_0)], \nonumber \\*    
\end{IEEEeqnarray} 
\else
\begin{IEEEeqnarray}{rCl}
	\label{Eq. BounCon2} 
	4\pi a^2 D_A \frac{\partial \pa{r}{t}}{\partial r} \bigg |_{r = a} & \hspace*{-1 mm} = & k_f \pa{a}{t} 	 
	 - k_b  [1 - S(t|r_0)],  
\end{IEEEeqnarray}
\fi
respectively, where $S(t|r_0)$ is the probability that a given $A$ molecule, released at distance $r_0$ at time $t_0$, is not reacting at the boundary of the receiver at time $t$ and has also not been degraded in the channel by time $t$. $S(t|r_0)$ can be obtained as follows: 
\begin{IEEEeqnarray}{c}
	\label{Eq. SurvivingProb} 
	S(t|r_0) = 1 - \int_{0}^{t} 4\pi a^2 D_A \frac{\partial \pa{r}{\tau}}{\partial r} \bigg |_{r = a} \dif \tau .  
\end{IEEEeqnarray}

The solution of reaction-diffusion equation (\ref{Eq. Reaction-Diffusion}) with initial and boundary conditions (\ref{Eq. InitialCondition})-(\ref{Eq. BounCon2}) is the Green's function\footnote{The solution of an inhomogeneous partial differential equation for an initial condition in the form of a Dirac delta function is referred to as the Green's function \cite{Ivar}.} of our system model. 

To summarize, obtaining the expected received signal $\overline{N}_C(t)$ requires knowledge of the channel impulse response $\pac{t}$ which, in turn, can be found via the Green's function based on (\ref{Eq. RelationshipReduced}).
\subsection{Dimensional Analysis} 
In this subsection, we provide a dimensional analysis and express our system model in dimensionless form. Dimensional analysis has the following advantages: 1) the generalization of our results is facilitated, 2) the round-off due to manipulations with large/small numbers is avoided, 3) the assessment of the relative importance of the parameters appearing in the equations in the following sections is facilitated, and 4) the dimensionless model reduces the number of parameters that appear in the equations. For the remainder of this paper, we transform all variables and equations into dimensionless form, where we denote the dimensionless variables by superscript ``$\prime$''. The corresponding dimensional variables can be obtained by scaling the dimensionless variables via reference variables.

Let us denote the reference distance in m and the reference number of molecules by $r_\text{ref}$ and $N_{A\text{ref}}$, respectively. Then, we can define the dimensionless time as $\di{t}=D_At/r_\text{ref}^2$ and dimensionless radial distance from the center of the receiver as $\di{r} = r/r_\text{ref}$. The dimensionless reaction rate constants in \eqref{Eq.DegReaction} and \eqref{Eq.RevReaction} can be written as       
\begin{IEEEeqnarray}{c}
	\label{Eq. DimensionlessReactionRates} 
		\dkd = \frac{\kd r_\text{ref}^2}{D_A}, \,\,\,\, \dkb = \frac{\kb r_\text{ref}^2}{D_A}, \,\,\,\, \dkf = \frac{\kf N_{A\text{ref}}}{D_A r_\text{ref}}.
\end{IEEEeqnarray} 

For consistency of notation, we also denote $\pa{r}{t}$ and $\pac{t}$ for dimensionless variables $\di{r}$, $\di{t}$, and $\di{r}_0$ as $\padi{r}{t}$ and $\pacdi{t}$, respectively. In the following, without loss of generality, we choose $r_\text{ref} = a$ and $N_{A\text{ref}} = 1$. Thus, \eqref{Eq. RelationshipReduced} in dimensionless form can be written as 
\begin{equation}
	\label{Eq. DiRelationshipReduced} 
	\pacdi{t} = \int_{0}^{\di{t}} 4\pi \frac{\partial \padi{r}{\tau}}{\partial \di{r}} \bigg |_{\di{r} = 1} \dif \di{\tau}.    
\end{equation} 

Furthermore, reaction-diffusion equation \eqref{Eq. Reaction-Diffusion} becomes 
\begin{equation}
	\label{Eq. DiReaction-Diffusion} 
	\frac{\partial \padi{r}{t}}{\partial \di{t}} = \nabla^{2} \padi{r}{t} - \dkd \padi{r}{t}, 
\end{equation} 
where
\ifdouble
\begin{IEEEeqnarray}{rCl}
	\frac{\partial \padi{r}{t}}{\partial \di{t}} & = & \frac{a^2}{D_A} \frac{\partial \pa{r}{t}}{\partial t}, \nonumber \\
	 \nabla^{2} \padi{r}{t} & = & a^2 \nabla^{2} \pa{r}{t}.  
\end{IEEEeqnarray}
\else 
\begin{equation}
	\frac{\partial \padi{r}{t}}{\partial \di{t}} = \frac{a^2}{D_A} \frac{\partial \pa{r}{t}}{\partial t}, \,\,\,\,\, \nabla^{2} \padi{r}{t} = a^2 \nabla^{2} \pa{r}{t}.  
\end{equation}
\fi

Initial condition \eqref{Eq. InitialCondition} and the first boundary condition \eqref{Eq. BounCon1} can be written as 
\begin{equation}
	\label{Eq. DiInitialCondition} 
	P_A(\di{r}, \di{t} \to 0 | \di{r}_0) = \frac{1}{4\pi {r_{0}^{\prime}}^{2}} \delta(\di{r} - \di{r}_0)
\end{equation} 
and 
\begin{equation}
	\label{Eq. DiBounCon1}
	\lim_{\di{r} \to \infty} \padi{r}{t} = 0,
\end{equation} 
respectively. The second boundary condition, given by \eqref{Eq. BounCon2} and \eqref{Eq. SurvivingProb}, simplifies in dimensionless form to 
\ifdouble
\begin{IEEEeqnarray}{rCl}
	\label{Eq. DiBounCon2} 
	4\pi \frac{\partial \padi{r}{t}}{\partial \di{r}} \bigg |_{\di{r} = 1} & = & \dkf P_A(1,\di{t}|\di{r}_0) \nonumber \\
	&& -\> \dkb \int_{0}^{\di{t}} \hs 4\pi \frac{\partial \padi{r}{\tau}}{\partial \di{r}} \bigg |_{\di{r} = 1} \dif \di{\tau}. \nonumber \\*      
\end{IEEEeqnarray}
\else 
\begin{IEEEeqnarray}{rCl}
	\label{Eq. DiBounCon2} 
	4\pi \frac{\partial \padi{r}{t}}{\partial \di{r}} \bigg |_{\di{r} = 1} & = & \dkf P_A(1,\di{t}|\di{r}_0) 	 
	 - \dkb \int_{0}^{\di{t}} 4\pi \frac{\partial \padi{r}{\tau}}{\partial \di{r}} \bigg |_{\di{r} = 1} \dif \di{\tau}.      
\end{IEEEeqnarray}
\fi 

\ifdouble
\newcounter{tempequationcounter} 
\begin{figure*}[!t]
\normalsize
\setcounter{tempequationcounter}{\value{equation}}
\begin{IEEEeqnarray}{rCl}
	\setcounter{equation}{23}
	\label{Eq. Green's function}
	\padi{r}{t} & = & \exp(-\dkd \di{t}) \Biggl[ \frac{1}{8 \pi \di{r} \di{r}_0 \sqrt{\pi \di{t}} } 
	 \left( \exp\left(\frac{-\left( \di{r} - \di{r}_0 \right)^2}{4 \di{t}}\right)  
	 + \exp\left( \frac{-\left( \di{r} + \di{r}_0 - 2 \right)^2}{4 \di{t}} \right) \right) - \frac{1}{4 \pi \di{r} \di{r}_0}  \nonumber \\
	&&  \times\>  \left( \di{\eta}_1 \, \w \left( \frac{\di{r} + \di{r}_0 - 2}{\sqrt{4\di{t}}}, \di{\alpha} \sqrt{\di{t}} \right) 
	 + \di{\eta}_2 \, \w \left( \frac{\di{r} + \di{r}_0 - 2}{\sqrt{4 \di{t}}}, \di{\beta} \sqrt{\di{t}}  \right)  
	  + \di{\eta}_3 \, \w \left( \frac{\di{r} + \di{r}_0 - 2}{\sqrt{4 \di{t}}}, \di{\gamma} \sqrt{\di{t}}  \right) \right)  \Biggr].
\end{IEEEeqnarray} 
\setcounter{equation}{\value{tempequationcounter}}
\hrulefill
\vspace*{4pt} 
\vspace*{-6 mm}
\end{figure*}
\fi
  
\subsection{Green's Function} 
In this subsection, we derive a closed-form analytical expression for the Green's function of the system. To this end, we adopt the methodology introduced in \cite{SchultenL1}. In particular, we decompose $\padi{r}{t}$ as 
\begin{IEEEeqnarray}{c}
	\label{Eq. Partition} 
	\padi{r}{t} = \di{U}(\di{r},\di{t}|\di{r}_0) + \di{V}(\di{r},\di{t} | \di{r}_0),
\end{IEEEeqnarray}
where function $\di{U}(\di{r},\di{t}|\di{r}_0)$ is chosen such that it satisfies both the reaction-diffusion equation (\ref{Eq. DiReaction-Diffusion}) and initial condition (\ref{Eq. DiInitialCondition}). On the other hand, function $\di{V}(\di{r},\di{t} | \di{r}_0)$ is chosen such that it satisfies (\ref{Eq. DiReaction-Diffusion}), but at the same time satisfies jointly with function $\di{U}(\di{r},\di{t} | \di{r}_0)$ boundary conditions (\ref{Eq. DiBounCon1}) and (\ref{Eq. DiBounCon2}). With this approach, we can decompose the original problem into two sub-problems as follows. In the first sub-problem, we solve the reaction-diffusion equation 
\begin{IEEEeqnarray}{c}
	\label{Eq. Reaction-DiffusionU} 
	\frac{\partial \di{U}(\di{r},\di{t}|\di{r}_0)}{\partial \di{t}} = \nabla^{2} \di{U}(\di{r},\di{t}|\di{r}_0) - \dkd \di{U}(\di{r},\di{t}|\di{r}_0), 
\end{IEEEeqnarray} 
with initial condition 
\begin{IEEEeqnarray}{c}
	\label{Eq. InitialConditionU} 
	\di{U}(\di{r},\di{t} \to 0 | \di{r}_0) = \frac{1}{4\pi {\di{r}_{0}}^{2}} \delta(\di{r} - \di{r}_0).
\end{IEEEeqnarray}
In the second sub-problem, we solve the reaction-diffusion equation
\begin{IEEEeqnarray}{c}
	\label{Eq. Reaction-DiffusionV} 
	\frac{\partial \di{V}(\di{r},\di{t} | \di{r}_0)}{\partial \di{t}} = \nabla^{2} \di{V}(\di{r},\di{t} | \di{r}_0) - \dkd \di{V}(\di{r},\di{t} | \di{r}_0), 
\end{IEEEeqnarray}
with the initial condition 
\begin{IEEEeqnarray}{c}
	\label{Eq. ConditionV} 
	\di{V}(\di{r},\di{t} \to 0 | \di{r}_0) = 0.
\end{IEEEeqnarray} 
Finally, the solutions of both sub-problems are combined, cf. (\ref{Eq. Partition}), such that they jointly satisfy boundary conditions (\ref{Eq. BounCon1}) and (\ref{Eq. BounCon2}). The final solution for the Green's function is given in the following theorem. 

\begin{theorem}[Green's function]
The probability of finding a given $A$ molecule at dimensionless distance $\di{r} \geq 1$ from the center of the receiver at dimensionless time $\di{t}$, given that it was released at dimensionless distance $\di{r}_0$ at dimensionless time $\di{t_0} = 0$ and may be degraded via first-order degradation reaction (\ref{Eq.DegReaction}) (with dimensionless reaction constant $\dkd$) and/or react with the receptor molecules at the receiver surface via second-order reversible reaction (\ref{Eq.RevReaction}) (with dimensionless forward reaction rate constant $\dkf$ and dimensionless backward reaction rate constant $\dkb$) during dimensionless time $\di{t}$ is given 
\ifdouble 
	by \eqref{Eq. Green's function}, shown at the top of the following page,
	\addtocounter{equation}{1} 
\else 
	as follows 
\fi 
where function $\w(n,m)$ is defined as 
\begin{IEEEeqnarray}{c} 
	\label{Eq. FunctionW} 
	\w(n,m) = \exp\left( 2nm + m^2 \right) \erfc\left( n + m \right),  
\end{IEEEeqnarray} 
$\erfc(\cdot)$ is the complementary error function, and constants $\di{\eta}_1$, $\di{\eta}_2$, and $\di{\eta}_3$ are given by 
\begin{IEEEeqnarray}{rCl} 
	\label{Eq. eta1} 
	\di{\eta}_1 & = & \frac{\di{\alpha}(\di{\gamma} + \di{\alpha})(\di{\alpha} + \di{\beta})}{(\di{\gamma} - \di{\alpha})(\di{\alpha} - \di{\beta})}, \\
	\label{Eq. eta2} 
	\di{\eta}_2 & = & \frac{\di{\beta}(\di{\gamma} + \di{\beta})(\di{\alpha} + \di{\beta})}{(\di{\beta} - \di{\gamma})(\di{\alpha} - \di{\beta})}, \\
	\label{Eq. eta3} 
	\di{\eta}_3 & = & \frac{\di{\gamma}(\di{\gamma} + \di{\beta})(\di{\alpha} + \di{\gamma})}{(\di{\beta} - \di{\gamma})(\di{\gamma} - \di{\alpha}
	)},
\end{IEEEeqnarray}
respectively. Here, $\di{\alpha}$, $\di{\beta}$, and $\di{\gamma}$ are the solutions of the following system of equations
\begin{equation}
	\left\{ \,
	\begin{IEEEeqnarraybox}[][c]{l?s}
	\IEEEstrut
	\di{\alpha} + \di{\beta} + \di{\gamma} = \left( 1 + \frac{\dkf}{4\pi} \right), \\
	\di{\alpha} \di{\gamma} + \di{\beta} \di{\gamma} + \di{\alpha} \di{\beta} = \dkb - \dkd, \\
	\di{\alpha} \di{\beta} \di{\gamma} = \dkb  - \dkd \left( 1 + \frac{\dkf}{4 \pi} \right).
	\IEEEstrut
	\end{IEEEeqnarraybox}
	\right.
	\label{Eq. AlphaBetaGamma}
\end{equation}           
\end{theorem} 

\begin{IEEEproof}
Please refer to the Appendix.
\end{IEEEproof}

\newtheorem{remark}{Remark}
\begin{remark} $\di{\alpha}$, $\di{\beta}$, and $\di{\gamma}$ may be complex numbers. As a result, the complex exponential and the complex complementary error function have to be used for evaluation of $\w(\cdot, \cdot)$. However, the sum of the three $\w(\cdot, \cdot)$ terms on the right hand side of (\ref{Eq. Green's function}) is always a real number. 
\end{remark} 


\subsection{Channel Impulse Response}
Given the Green's function derived in the previous section, we can calculate the channel impulse response via (\ref{Eq. DiRelationshipReduced}). This leads to 
\ifdouble 
\begin{IEEEeqnarray}{rCl} 
	\label{Eq. Channel_Impulse_Response}
 \pacdi{t} & = & \frac{\di{k}_f e^{-\di{k}_d \di{t}}}{4\pi \di{r}_0} \left\lbrace \frac{\di{\alpha} \w\left( \frac{\di{r}_0 - 1}{\sqrt{ 4 \di{t}}}, \di{\alpha} \sqrt{\di{t}} \right)}{(\di{\gamma} - \di{\alpha})(\di{\alpha} - \di{\beta})}  \right. \nonumber \\
 && \hspace*{-2 mm} \left. +\> \frac{\di{\beta}  \w\left( \frac{\di{r}_0 - 1}{\sqrt{4 \di{t}}}, \di{\beta} \sqrt{\di{t}} \right)}{(\di{\beta} - \di{\gamma})(\di{\alpha} - \di{\beta})}  
  + \hspace*{-1 mm} \frac{\di{\gamma} \w\left( \frac{\di{r}_0 - 1}{\sqrt{4 \di{t}}}, \di{\gamma} \sqrt{\di{t}} \right)}{(\di{\beta} - \di{\gamma})(\di{\gamma} - \di{\alpha})} \right\rbrace. \nonumber \\*  	
\end{IEEEeqnarray} 
\else 
\begin{IEEEeqnarray}{rCl} 
	\label{Eq. Channel_Impulse_Response}
 \pacdi{t} & = & \frac{\di{k}_f e^{-\di{k}_d \di{t}}}{4\pi \di{r}_0} \left\lbrace \frac{\di{\alpha} \w\left( \frac{\di{r}_0 - 1}{\sqrt{ 4 \di{t}}}, \di{\alpha} \sqrt{\di{t}} \right)}{(\di{\gamma} - \di{\alpha})(\di{\alpha} - \di{\beta})}  
  + \frac{\di{\beta}  \w\left( \frac{\di{r}_0 - 1}{\sqrt{4 \di{t}}}, \di{\beta} \sqrt{\di{t}} \right)}{(\di{\beta} - \di{\gamma})(\di{\alpha} - \di{\beta})}  
  + \frac{\di{\gamma} \w\left( \frac{\di{r}_0 - 1}{\sqrt{4 \di{t}}}, \di{\gamma} \sqrt{\di{t}} \right)}{(\di{\beta} - \di{\gamma})(\di{\gamma} - \di{\alpha})} \right\rbrace. \hspace*{7 mm}   	
\end{IEEEeqnarray}
\fi 

Finally, the dimensionless number of $C$ molecules expected on the surface of the receiver after impulsive release of  a dimensionless number $\di{N}_A$ of $A$ molecules at the transmitter can be obtained as 
\begin{IEEEeqnarray}{c} 
	\label{Eq. ExpectedReceSignal} 
	\di{\overline{N}}_C(t) = \di{N}_A \pacdi{t}.
\end{IEEEeqnarray} 

In the remainder of this subsection, we consider the scenarios studied in \cite{YilmazL1} and \cite{Heren}, respectively, and show that the results provided there, are special cases of \eqref{Eq. Channel_Impulse_Response}. In particular, for the first scenario, similar to \cite{Heren}, we assume that the reaction on the boundary of the receiver is \emph{irreversible} (i.e., $\dkb = 0$), every collision of an information molecule with the receiver surface leads to the formation of a $C$ molecule (i.e., $\dkf \to \infty$), and there are degradation reactions in the channel (i.e., $\dkd \neq 0$). In the second scenario, similar to \cite{YilmazL1}, we assume that there are no first-order degradation reactions in the channel (i.e., $\dkd = 0$), while $\dkf \to \infty$ and $\dkb = 0$. 

\begin{corollary}[Irreversible reaction with degradation]
For the case of irreversible reaction with degradation, i.e., $\dkb = 0$, $\dkf \to \infty$, and $\dkd > 0$, the dimensionless channel impulse response \eqref{Eq. Channel_Impulse_Response} simplifies to
\ifdouble
\begin{IEEEeqnarray}{rcl} 
	\label{Eq. corollary2} 
	 \pacdi{t} & = & \frac{1}{2 \di{r}_0} \left\lbrace \exp \left( \sqrt{\dkd} (\di{r}_0 -1) \right)  \right. \nonumber \\
	 && \left. \times\> \erfc \hs \left( \frac{\di{r}_0 \hs - \hs 1}{\sqrt{4\di{t}}} \hs + \hs \sqrt{\dkd \di{t}} \right)
	 \hs +  \exp \left( \hs - \sqrt{\dkd} (\di{r}_0 \hs - \hs 1) \hs \right) \right. \nonumber \\ 
	&& \left. \times \>  \erfc \left( \frac{\di{r}_0 -1}{\sqrt{4\di{t}}} - \sqrt{\dkd \di{t}} \right) \right\rbrace. 
\end{IEEEeqnarray}
\else 
\begin{IEEEeqnarray}{rcl} 
	\label{Eq. corollary2} 
	 \pacdi{t} & = & \frac{1}{2 \di{r}_0} \left\lbrace \exp \left( \sqrt{\dkd} (\di{r}_0 -1) \right) \erfc \left( \frac{\di{r}_0 -1}{\sqrt{4\di{t}}} + \sqrt{\dkd \di{t}} \right) \right. \nonumber \\ 
	 && \left. +\> \exp \left( - \sqrt{\dkd} (\di{r}_0 -1) \right) \erfc \left( \frac{\di{r}_0 -1}{\sqrt{4\di{t}}} - \sqrt{\dkd \di{t}} \right) \right\rbrace.
\end{IEEEeqnarray}
\fi 
\begin{IEEEproof} 
Let us first consider the case where $\dkb = 0$, and $\dkd$ and $\dkf$ are finite numbers. Then, it is straightforward to show that the system of equations in \eqref{Eq. AlphaBetaGamma} has the following solutions
\begin{IEEEeqnarray}{rcl} 
	\label{Eq. corollary2SolAlphaBetaGamma}  
	\di{\alpha} = \left( 1 + \frac{\dkf}{4\pi} \right), \,\,\,\, \di{\beta} = \sqrt{\dkd}, \,\,\,\, \di{\gamma} = -\sqrt{\dkd}.
\end{IEEEeqnarray} 
Substituting the values of $\di{\alpha}$, $\di{\beta}$, and $\di{\gamma}$ from \eqref{Eq. corollary2SolAlphaBetaGamma} into \eqref{Eq. Channel_Impulse_Response} and considering that in the limit of $\dkf \to \infty$, $\di{\alpha} \to \infty$, and as a result of this, the first $\w(\cdot, \cdot)$ on the right hand side of \eqref{Eq. Channel_Impulse_Response} becomes zero, \eqref{Eq. Channel_Impulse_Response} simplifies to \eqref{Eq. corollary2}.     
\end{IEEEproof} 

It can be easily verified that the dimensional form of \eqref{Eq. corollary2} is identical to \cite[Eq. (12)]{Heren}.
\end{corollary} 

\begin{corollary}[Irreversible reaction without degradation]
For the case of irreversible reaction without degradation, i.e., $\dkb = 0$, $\dkf \to \infty$, and $\dkd = 0$, the dimensionless channel impulse response \eqref{Eq. Channel_Impulse_Response} simplifies to 
\begin{IEEEeqnarray}{rcl}
	\label{Eq. corollary1} 
	\pacdi{t} & = & \frac{1}{\di{r}_0} \erfc \left( \frac{\di{r}_0 - 1}{\sqrt{4 \di{t}}} \right).
\end{IEEEeqnarray}
\end{corollary}

\begin{IEEEproof}
It can be easily verified that \eqref{Eq. corollary1} is obtained from \eqref{Eq. corollary2} after setting $\dkd = 0$.  
\end{IEEEproof}
The dimensional form of \eqref{Eq. corollary1} is identical to \cite[Eq. (23)]{YilmazL1}. However, this represents only the special case of an irreversible reaction on the surface of the receiver with $\dkf \to \infty$. 
\section{expected received signal for finite number of receptors} 
\label{Sec. FiniteRecNum} 
In this section, we study the impact of individual receptor modeling on the channel impulse response derived in the previous section. Thereby, we take into account that the surface of the receiver is only partially covered with receptor protein molecules. 
\subsection{Problem Formulation and Some Preliminaries} 
In order to study the impact of individual receptor modeling, we model each receptor protein molecule as a circle with radius $r_s$ mounted on the surface of the receiver. We assume that a total of $M$ receptors are \emph{uniformly} distributed and partially cover the surface of the receiver, where the forward and backward reaction rate constants of each receptor are $\kf$ and $\kb$, respectively. With these assumptions, the incoming probability flux, $-J(\vec{r},t | \vec{r}_0)$, at the surface of the receiver can be written in dimensional form as 
\ifdouble
\begin{equation}
	\left\{ \,
	\begin{IEEEeqnarraybox}[][c]{l?s}
	\IEEEstrut
	\hs \hs -J(\vec{r},t | \vec{r}_0)\big|_{\vec{r} \in \Omega_s} \hs = D_A \nabla \pav{\vec{r}}{t}\big|_{\vec{r} \in \Omega_s}  \text{on the receptor}, \\
	\hs \hs \hspace{0.7 mm} - \hspace{0.9 mm} J(\vec{r},t | \vec{r}_0)\big|_{\vec{r} \in \Omega_r} \hs = 0 \hspace{3 cm}\text{off the receptor},
	\IEEEstrut
	\end{IEEEeqnarraybox}
	\right.
	\label{Eq. probfluxReceptors}
\end{equation}
\else 
\begin{equation}
	\left\{ \,
	\begin{IEEEeqnarraybox}[][c]{l?s}
	\IEEEstrut
	-J(\vec{r},t | \vec{r}_0)\big|_{\vec{r} \in \Omega_s} = D_A \nabla \pav{\vec{r}}{t}\big|_{\vec{r} \in \Omega_s} \hspace{2 cm} \text{on the receptor}, \\
	-J(\vec{r},t | \vec{r}_0)\big|_{\vec{r} \in \Omega_r} \hspace{.2 cm} = 0 \hspace{5.4cm} \text{off the receptor},
	\IEEEstrut
	\end{IEEEeqnarraybox}
	\right.
	\label{Eq. probfluxReceptors}
\end{equation}
\fi
where $\Omega_s$ is the surface of the $i$th receptor, $i \in \{1, \cdots, M \}$, and $\Omega_r$ is the part of the receiver surface that is not covered by receptors. Since we assume now that the surface of the receiver is only partially covered by type $B$ molecules, i.e., receptors, $J(\vec{r},t | \vec{r}_0)$ and $\pacv{t}$ are functions of $r$, $\theta$, and $\phi$ on the surface of the receiver. This makes the evaluation of \eqref{Eq. Relationship} very difficult. Furthermore, modeling the second boundary condition for the reaction mechanism in \eqref{Eq.RevReaction} as in \eqref{Eq. BounCon2} is no longer possible. Instead, each individual receptor has its own boundary condition on the surface of the receiver, which makes the overall boundary condition for the entire receiver surface \emph{heterogeneous}, and, as a result, the analysis becomes intractable. In order to overcome this problem, we employ a technique which is referred to as boundary homogenization, see \cite{Berg1, Zwanzig1, Zwanzig2, Berezhkovskii1, Berezhkovskii2}. 

Specifically, when applying boundary homogenization, a heterogeneous boundary condition is approximated by a homogeneous one. In particular, for the problem at hand, a receiver whose surface is partially covered with receptor $B$ molecules having dimensional forward reaction rate $\kf$ is approximated by a receiver whose surface is \emph{fully} covered with receptor $B$ molecules having a \emph{modified} forward reaction rate, which we denote by $\kfm$. 

Let us denote the total steady-state flux of $A$ molecules into a receiver fully covered by receptor $B$ molecules by $\js$ and the total steady-sate flux of $A$ molecules into a receiver partially covered by $M$ receptor $B$ molecules by $\jsmr$. Furthermore, we represent the total steady-state flux of $A$ molecules into a given receptor by $\jr$. In their pioneering work on the theory of steady-state ligand-receptor binding \cite{Berg1}, Berg and Purcell applied the steady-state theory of ligand-receptor binding to a sphere partially covered by receptors. They could show that for the case when the receptors are fully absorbing, $\jsmr$ can be approximated as \cite[Eq. (8)]{Berg1} 
\ifdouble
\begin{IEEEeqnarray}{rCl} 
	\label{Eq. BurgPurcell} 
	\jsmr & = & \js \times \overbrace{\frac{M \jr}{\js + M \jr}}^\text{Correction factor $\varphi$} 
	 =  4 \pi D_A C_{\infty} \times \frac{M r_s}{\pi a + M r_s}, \nonumber \\*  
\end{IEEEeqnarray}
\else
\begin{IEEEeqnarray}{rCl} 
	\label{Eq. BurgPurcell} 
	\jsmr & = & \js \times \overbrace{\frac{M \jr}{\js + M \jr}}^\text{Correction factor $\varphi$} 
	 =  4 \pi D_A C_{\infty} \times \frac{M r_s}{\pi a + M r_s},  
\end{IEEEeqnarray} 
\fi
where $\js = 4 \pi D_A C_{\infty}$ for a fully absorbing receiver, i.e., $\kf \to \infty$, and $\jr = 4 D_A r_s$ for a fully absorbing circular receptor. $C_{\infty}$ is the initial concentration of $A$ molecules outside the receiver, where we assume that $C_{\infty} = 1$ to be consistent with initial boundary condition \eqref{Eq. InitialCondition}. Eq. \eqref{Eq. BurgPurcell} indicates that for given $\js$ and $\jr$ one can approximate $\jsmr$ as $\jsmr = \js \times \varphi$, where $\varphi$ is referred to as the correction factor. Correction factor $\varphi$ is also used in \cite{Akkaya} to approximate the expected received signal for a receiver partially covered with fully absorbing receptors, see \cite[Eq. (10)]{Akkaya}. 

The result of Berg and Purcell was re-derived by Zwanzig \cite{Zwanzig1}, where a modification factor $\lambda$ was introduced in the denominator of \eqref{Eq. BurgPurcell} as follows \cite[Eq. (5)]{Zwanzig1} 
\begin{IEEEeqnarray}{rCl} 
	\label{Eq. ZwanzigFactor} 
	\jsmr & = & \js \times \overbrace{\frac{M \jr}{(1 - \lambda)\js + M \jr}}^\text{Correction factor $\varphi$},  
\end{IEEEeqnarray} 
where $\lambda$ is the fraction of receiver surface covered by receptors, i.e., $\lambda = M \frac{\pi r_s^2}{4 \pi a^2}$. The results in both \cite{Berg1} and \cite{Zwanzig1} were derived under the assumptions that 1) the surface area of each receptor is small compared to the surface area of the receiver so that the effect of receptor occupancy is negligible and 2) the distance between any pair of receptors is large enough so that the flux $\jr$ for a given receptor is independent of the flux $\jr$ of all other receptors. In the following, we also adopt these assumptions. Furthermore, the analyses in \cite{Berg1} and \cite{Zwanzig1} require the assumption that the receptors are fully absorbing. The authors in \cite{Zwanzig2} showed that \eqref{Eq. ZwanzigFactor} can also be employed for evaluation of $\jsmr$ when the receptors are partially absorbing, where $\js$ for a partially absorbing receiver and $\jr$ for a partially absorbing receptor are given by \cite[Eq. (13)]{Zwanzig2} 
\begin{IEEEeqnarray}{C} 
	\label{Eq. ZwanzigParAbsRec} 
	\js =  \frac{4 \pi \kf D_A a^2}{a \kf + 4 \pi D_A},
\end{IEEEeqnarray} 
and \cite[Eq. (15)]{Zwanzig2}
\begin{IEEEeqnarray}{C} 
	\label{Eq. ZwanzigParAbsRes} 
	\jr =  \frac{4 r_s D_A}{1 + 16D_A/(r_s\kf)},
\end{IEEEeqnarray} 
respectively. Now, given $\js$ and $\jr$ in \eqref{Eq. ZwanzigParAbsRec} and \eqref{Eq. ZwanzigParAbsRes}, correction factor $\varphi$ can be obtained via \eqref{Eq. ZwanzigFactor} as follows
\ifdouble
\begin{IEEEeqnarray}{C} 
	\label{Eq. CorrectionFactor} 
	\varphi = \frac{M r_s^2 (\kf a + 4 \pi D_A)}{a^2(1 - \lambda)(\pi r_s \kf + 16 \pi D_A) + M r_s^2 (\kf a + 4 \pi D_A)}. \nonumber \\*
\end{IEEEeqnarray} 
\else
\begin{IEEEeqnarray}{C} 
	\label{Eq. CorrectionFactor} 
	\varphi = \frac{M r_s^2 (\kf a + 4 \pi D_A)}{a^2(1 - \lambda)(\pi r_s \kf + 16 \pi D_A) + M r_s^2 (\kf a + 4 \pi D_A)}. 
\end{IEEEeqnarray} 
\fi

\subsection{Channel Impulse Response} 
In order to derive the channel impulse response for a receiver that is partially covered by $M$ receptors, which we denote by $\pacm{t}$, we employ the boundary homogenization technique similar to \cite{Berezhkovskii1} and \cite{Berezhkovskii2}. In particular, we assume that there is a receiver that is uniformly covered by receptor $B$ molecules with modified forward reaction rate $\kfm$ such that $\pacm{t} = \pacn{t}$, where $\pacn{t}$ is the channel impulse response of the receiver uniformly covered by receptor molecules with modified forward reaction rate $\kfm$. For deriving an analytical expression that relates $\kfm$ to the other system parameters, we consider the steady-state regime, i.e., we assume that $\pacm{t \to \infty} = \pacn{t \to \infty}$. To this end, given \eqref{Eq. ZwanzigFactor} and \eqref{Eq. CorrectionFactor}, the relationship between $\pacm{t \to \infty}$ and $\pac{t \to \infty}$ can be expressed as follows 
\begin{IEEEeqnarray}{C} 
	\label{Eq. PacmInfinityRatio} 
	\frac{\pacm{t \to \infty}}{\pac{t \to \infty}} = \frac{\jsmr}{\js} = \varphi , 
\end{IEEEeqnarray}         
where we used the fact that the \emph{ratio} $\pacm{t} / \pac{t}$ in the limit of $t \to \infty$ is equal to the ratio $\jsmr / \js$. Now, given \eqref{Eq. PacmInfinityRatio}, we can write
\ifdouble
\begin{IEEEeqnarray}{C} 
	\label{Eq. PacmRelationshipToPacModified} 
	\pacm{t \to \infty} = \pac{t \to \infty} \times \varphi = \pacn{t \to \infty}. \nonumber \\*
\end{IEEEeqnarray}
\else
\begin{IEEEeqnarray}{C} 
	\label{Eq. PacmRelationshipToPacModified} 
	\pacm{t \to \infty} = \pac{t \to \infty} \times \varphi = \pacn{t \to \infty}.
\end{IEEEeqnarray}
\fi
 
Finally, we have to find the asymptotic value of $\pac{t \to \infty}$ (and $\pacn{t \to \infty}$). However, it can be easily verified from the dimensional form of \eqref{Eq. Channel_Impulse_Response} that when $\kb > 0$ and/or $\kd > 0$, $\pac{t \to \infty} = 0$ (and $\pacn{t \to \infty} = 0$). This can also be verified intuitively, since when $\kb > 0$ and/or $\kd > 0$, the ultimate fate of a given information $A$ molecule is either dissociation or degradation in the limit $t \to \infty$. This provides a trivial solution for \eqref{Eq. PacmRelationshipToPacModified}, i.e., $0 \times \varphi = 0$. To overcome this problem, we consider in the following corollary $\pac{t \to \infty}$ when $\kb = \kd = 0$ and $\kf \neq 0$. 
\begin{corollary}
The dimensional form of the channel impulse response in \eqref{Eq. Channel_Impulse_Response} in the limit of $t \to \infty$, when $\kb = \kd = 0$ and $\kf \neq 0$, simplifies to 
	\begin{IEEEeqnarray}{C} 
		\label{Eq. PacAsymptotic} 
		\pac{t \to \infty} = \frac{a}{r_0} \times \frac{\kf a}{ \kf a + 4 \pi D_A}. 
	\end{IEEEeqnarray}     
\end{corollary} 

\begin{IEEEproof}
When $\dkb = \dkd = 0$ and $\dkf \neq 0$, it can be easily verified that the system of equations in \eqref{Eq. AlphaBetaGamma} has the following solutions 
\begin{IEEEeqnarray}{rcl} 
	\label{Eq. corollary22SolAlphaBetaGamma}  
	\di{\alpha} = \left( 1 + \frac{\dkf}{4\pi} \right), \,\,\,\, \di{\beta} = 0, \,\,\,\, \di{\gamma} = 0.
\end{IEEEeqnarray} 
Substituting the values of $\di{\beta}$ and $\di{\gamma}$ into \eqref{Eq. Channel_Impulse_Response} leads to 
\ifdouble
\begin{IEEEeqnarray}{rCl} 
	\label{Eq. corollary22}
 \pacdi{t} & = & \frac{\di{k}_f}{4\pi r_0} \left\lbrace \hs \frac{-\w\left( \frac{\di{r}_0 - 1}{\sqrt{ 4 \di{t}}}, \di{\alpha} \sqrt{\di{t}} \right)}{\di{\alpha}}  
  + \frac{\w\left( \frac{\di{r}_0 - 1}{\sqrt{4 \di{t}}}, 0 \right)}{\di{\alpha}}  
   \right\rbrace. \nonumber \\* 	
\end{IEEEeqnarray}
\else
\begin{IEEEeqnarray}{rCl} 
	\label{Eq. corollary22}
 \pacdi{t} & = & \frac{\di{k}_f}{4\pi r_0} \left\lbrace \frac{-\w\left( \frac{\di{r}_0 - 1}{\sqrt{ 4 \di{t}}}, \di{\alpha} \sqrt{\di{t}} \right)}{\di{\alpha}}  
  + \frac{\w\left( \frac{\di{r}_0 - 1}{\sqrt{4 \di{t}}}, 0 \right)}{\di{\alpha}}  
   \right\rbrace.  	
\end{IEEEeqnarray}
\fi
Now, substituting the value of $\di{\alpha}$ from \eqref{Eq. corollary22SolAlphaBetaGamma} into \eqref{Eq. corollary22}  and considering that $\w(n,m)|_{m=0} = \erfc(n)$ (see \eqref{Eq. FunctionW}), \eqref{Eq. corollary22} can be written as
\ifdouble
\begin{IEEEeqnarray}{rCl} 
	\label{Eq. corollary23}
 \pacdi{t} & = & \frac{1}{\di{r}_0} \times \frac{\dkf}{\dkf + 4 \pi} \left\lbrace -\w\left( \frac{\di{r}_0 - 1}{\sqrt{ 4 \di{t}}}, \di{\alpha} \sqrt{\di{t}} \right) \right. \nonumber \\
 && \left.  +\> \erfc \left( \frac{\di{r}_0 - 1}{\sqrt{4 \di{t}}} \right)  
   \right\rbrace.  	
\end{IEEEeqnarray}
\else
\begin{IEEEeqnarray}{rCl} 
	\label{Eq. corollary23}
 \pacdi{t} & = & \frac{1}{\di{r}_0} \times \frac{\dkf}{\dkf + 4 \pi} \left\lbrace -\w\left( \frac{\di{r}_0 - 1}{\sqrt{ 4 \di{t}}}, \di{\alpha} \sqrt{\di{t}} \right) 
  + \erfc \left( \frac{\di{r}_0 - 1}{\sqrt{4 \di{t}}} \right)  
   \right\rbrace.  	
\end{IEEEeqnarray}
\fi 
In the limit of $\di{t} \to \infty$, the functions $\w(\cdot, \cdot)$ and $\erfc(\cdot)$ become zero and one, respectively, and $\pacdi{t}$ approaches the dimensionless asymptotic value $\frac{1}{\di{r}_0} \times \frac{\dkf}{\dkf + 4 \pi} $. Transforming this asymptotic value into dimensional form leads to \eqref{Eq. PacAsymptotic}. 
\end{IEEEproof}

Eq. \eqref{Eq. PacAsymptotic} is also valid for $\pacn{t \to \infty}$ after substituting $\kf$ with $\kfm$. Now, given \eqref{Eq. PacAsymptotic}, \eqref{Eq. PacmRelationshipToPacModified} can be rewritten as 
\begin{IEEEeqnarray}{C}
	\label{Eq. Eq. PacmRelationshipToPacModifiedReduced} 
	\frac{a}{r_0} \times \frac{\kf a}{ \kf a + 4 \pi D_A} \times \varphi = \frac{a}{r_0} \times \frac{\kfm a}{ \kfm a + 4 \pi D_A}.
\end{IEEEeqnarray} 
Solving the above equation for $\kfm$ yields the final expression describing the relationship between $\kfm$ and the other system parameters as follows 
\begin{IEEEeqnarray}{C}
	\label{Eq. KfModifiedFinalRelasionship} 
	\kfm = \frac{4 \pi D_A \kf \varphi}{\kf a (1 - \varphi) + 4 \pi D_A},
\end{IEEEeqnarray}
where $\varphi$ is given by \eqref{Eq. CorrectionFactor}. Eq. \eqref{Eq. KfModifiedFinalRelasionship} can be written in  dimensionless form as follows 
\begin{IEEEeqnarray}{C}
	\label{Eq. DiKfModifiedFinalRelasionship} 
	\dkfm = \frac{4 \pi \dkf \di{\varphi}}{\dkf (1 - \di{\varphi}) + 4 \pi},
\end{IEEEeqnarray}  
where dimensionless $\di{\varphi}$ can be evaluated as 
\begin{IEEEeqnarray}{C} 
	\label{Eq. DiCorrectionFactor} 
	\di{\varphi} = \frac{M {r}^{\prime 2}_s (\dkf + 4 \pi)}{(1 - \di{\lambda})(\pi \di{r}_s \dkf + 16 \pi) + M {r}^{\prime 2}_s (\dkf + 4 \pi)},  
\end{IEEEeqnarray}
here $\di{\lambda} = M \frac{\pi {r}^{\prime 2}_s}{4 \pi}$.
  
To summarize, for evaluation of the dimensionless channel impulse response of a receiver partially covered by $M$ receptors, ${P}^{\prime M}_{AC}(\di{t}| \di{r}_0)$, having dimensionless reaction rates $\dkf$, $\dkb$, and $\dkd$, we first solve the system of equation in \eqref{Eq. AlphaBetaGamma} after substituting $\dkf$ with $\dkfm$. Then, after substituting $\dkf$ with $\dkfm$ and finding the corresponding values of $\di{\alpha}$, $\di{\beta}$, and $\di{\gamma}$, ${P}^{\prime M}_{AC}(\di{t}| \di{r}_0)$ can be evaluated via \eqref{Eq. Channel_Impulse_Response}. We assess the accuracy of the proposed approximate channel impulse response in Section \ref{Sec. Simulations}.          
\section{simulation framework}
\label{Sec. SimulationFramework} 
In this section, we provide a detailed description of the simulation framework developed to assess the accuracy of the analytical expressions proposed in Sections \ref{Sec.CIR} and \ref{Sec. FiniteRecNum}. 

We perform a Brownian motion particle-based simulation, where the precise locations of all individual molecules are
tracked throughout the simulation environment. In the adopted simulation algorithm, time is advanced in discrete steps of $\Delta t$ seconds. In order to jointly simulate the reactions, i.e., (\ref{Eq.DegReaction}) and (\ref{Eq.RevReaction}), and the diffusion of the molecules, we combine in our simulator the algorithm proposed for the simulation of first-order reactions in \cite{Steven_Andrews} with the algorithm for simulation of second-order reversible reactions introduced in \cite{WoldeJ1}. Furthermore, we adopt a triangle mesh grid for constructing the surface of the receiver, see Fig. \ref{Fig.SimulationModel}. Prior to simulation, $M$ triangles, which are shown in blue color in Fig. \ref{Fig.SimulationModel}, are chosen randomly out of all triangles constructing the surface of the receiver and represent the set of $M$ receptors. In particular, in each step of the simulation, we perform the following operations:
\ifdouble
\begin{figure}[!t] 
	\centering
	\includegraphics[scale = 0.6]{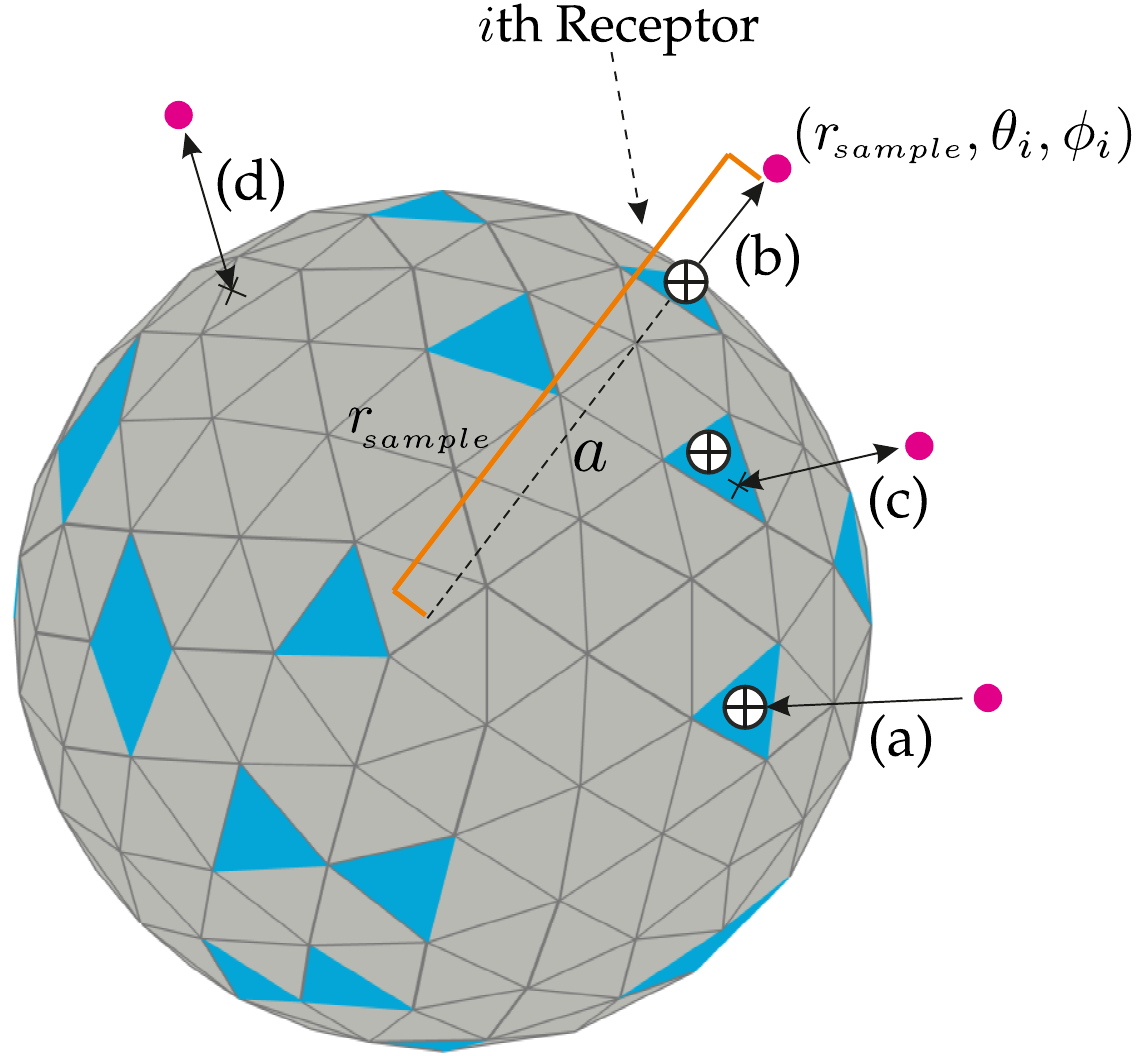}
	\caption{Schematic diagram of the considered triangle mesh grid for constructing the receiver, where the receptor and non-receptor parts of the receiver surface are shown with blue and gray triangles, respectively. The information $A$ molecules and activated receptor $C$ molecules are depicted as red circle and white circle with a ``+'' inside, respectively. One sample displacement of an $A$ molecule that leads to the production of a $C$ molecule is shown as an arrow marked with (a). The production of an $A$ molecule after occurrence of a backward reaction is shown as an arrow marked with (b). The reflections of an $A$ molecule upon contact with an occupied receptor and the part of the receiver that is not covered by receptors are shown with double arrows marked with (c) and (d), respectively.} 
	\label{Fig.SimulationModel}
\end{figure}
\else
\begin{figure}[!t] 
	\centering
	\includegraphics[scale = 0.6]{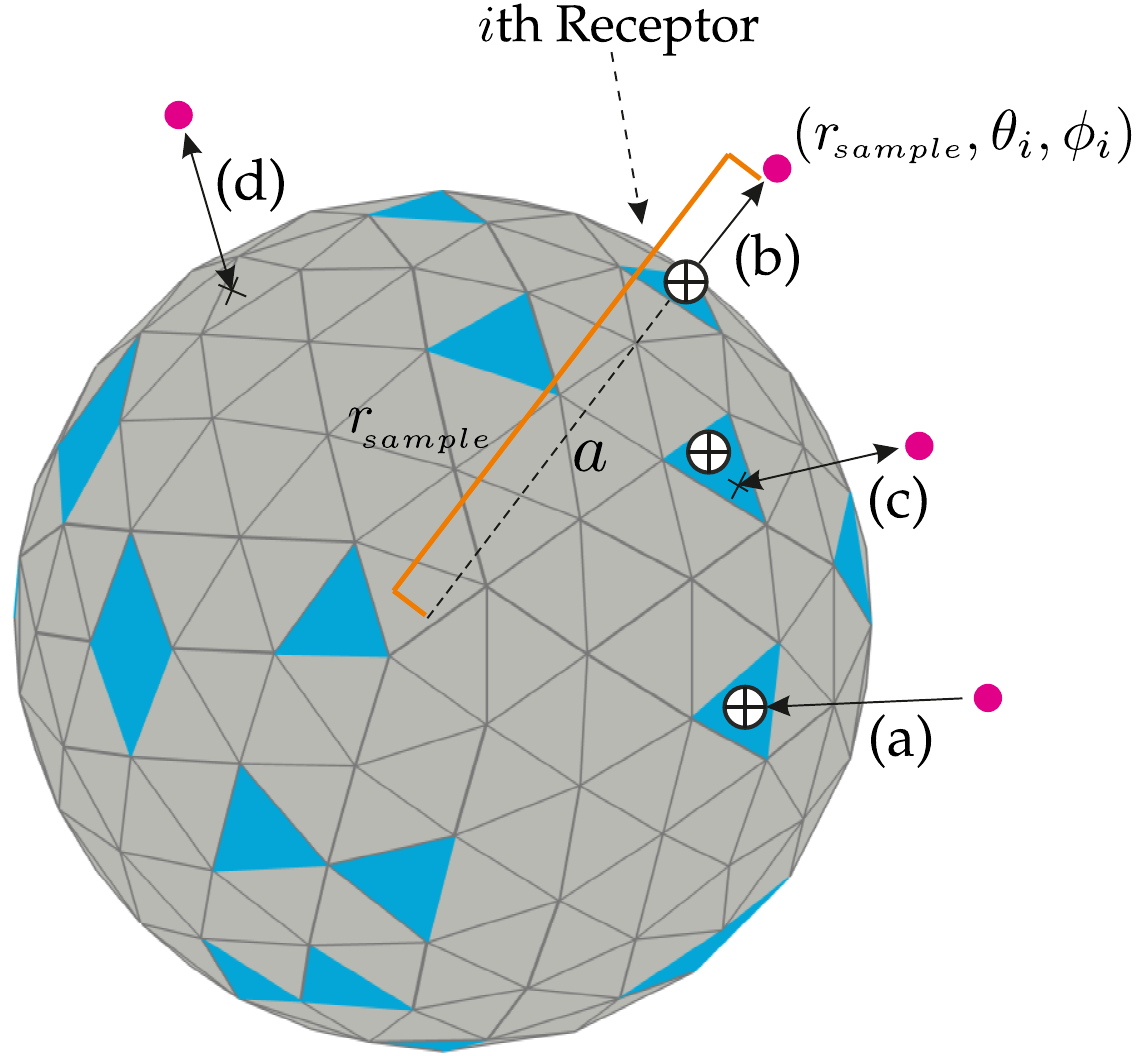}\vspace*{-2 mm}
	\caption{Schematic diagram of the considered triangle mesh grid for constructing the receiver, where the receptor and non-receptor parts of the receiver surface are shown with blue and gray triangles, respectively. The information $A$ molecules and activated receptor $C$ molecules are depicted as red circle and white circle with a ``+'' inside, respectively. One sample displacement of an $A$ molecule that leads to the production of a $C$ molecule is shown as an arrow marked with (a). The production of an $A$ molecule after occurrence of a backward reaction is shown as an arrow marked with (b). The reflections of an $A$ molecule upon contact with an occupied receptor and the part of the receiver that is not covered by receptors are shown with double arrows marked with (c) and (d), respectively.}\vspace*{-5 mm} 
	\label{Fig.SimulationModel}
\end{figure}
\fi
 
$1$) Each $A$ molecule undergoes a random walk, where the new position of the molecule in each Cartesian coordinate is obtained by sampling a Gaussian random variable with mean $0$ and variance $\sqrt{2D_A\Delta t}$.

$2$) A uniformly distributed random number $l_1 \in [0,1]$ is generated for each $A$ molecule. Then, a given $A$ molecule degrades and is removed from the environment if its $l_1 \leq \pr \left(\text{Reaction} \, k_d \right)$, where $\pr \left(\text{Reaction} \, k_d \right)$ is the degradation probability of a given $A$ molecule in $\Delta t$ seconds, and is given by \cite[Eq. (13)]{Steven_Andrews} 
\begin{IEEEeqnarray}{c} 
	\label{Eq. ProbUniMolReac} 
	\pr \left(\text{Reaction} \, k_d \right) = 1 - \exp(-k_d \Delta t).
\end{IEEEeqnarray}

$3$) If the final position of an $A$ molecule at the end of a simulation step leads to an overlap with the $i$th receptor, $i \in \{1, \cdots, M \}$, a uniformly distributed random number $l_2 \in [0,1]$ is generated. Then, this displacement is accepted as the occurrence of a forward reaction if $l_2 \leq \pr \left( \text{Reaction} \, k_f \right)$. $\pr \left( \text{Reaction} \, k_f \right)$ is the probability of the forward reaction and given by \cite[Eq. (22)]{WoldeJ1} 
\begin{IEEEeqnarray}{c}
	\label{Eq. ProbBiMolReacForwardAcc} 
	\pr \left( \text{Reaction} \, k_f \right) = \frac{k_f \Delta t}{4\pi \rho},
\end{IEEEeqnarray}
where $\rho$ is a normalization factor that can be evaluated as 
\begin{IEEEeqnarray}{c} 
	\label{Eq. ForwardReacNormFac} 
	\rho = \int_{a}^{\infty} \Pr \left( \text{Ovr} | \vec{r}, \Delta t \right) r^2 \dif r.
\end{IEEEeqnarray}
Here, $\Pr \left( \text{Ovr} | \vec{r}, \Delta t \right)$ is the probability that a given $A$ molecule at position $\vec{r}$ \emph{overlaps} with the \emph{receiver surface} in $\Delta t$ seconds, and can be written as \cite[Eq. (B3)]{WoldeJ1} 
\ifdouble
\begin{IEEEeqnarray}{rCl}
	\label{Eq. ProbOverlap} 
	\Pr \left( \text{Ovr} | \vec{r}, \Delta t \right) & = & \frac{a}{2r \sqrt{\pi}} \left[ \exp\left(\frac{-(r+a)^2}{\sigma^2}\right) \right. \nonumber \\ 
	&& \hspace*{-27 mm} \left. -\>  \exp\left( \frac{-(r-a)^2}{\sigma^2} \right) \right] \hs + \hs \frac{1}{2} \left[ \erf\left( \frac{r+a}{\sigma} \right)  
	  + \erf\left( \frac{a-r}{\sigma} \right) \right], \nonumber \\* 
\end{IEEEeqnarray} 
\else 
\begin{IEEEeqnarray}{rCl}
	\label{Eq. ProbOverlap} 
	\Pr \left( \text{Ovr} | \vec{r}, \Delta t \right) & = & \frac{a}{2r \sqrt{\pi}} \left[ \exp\left(\frac{-(r+a)^2}{\sigma^2}\right) - \exp\left( \frac{-(r-a)^2}{\sigma^2} \right) \right] \nonumber \\ 
	&&  +\> \frac{1}{2} \left[ \erf\left( \frac{r+a}{\sigma} \right)  
	  + \erf\left( \frac{a-r}{\sigma} \right) \right],
\end{IEEEeqnarray} 
\fi
where $\sigma^2 = 4 D_A \Delta t$ and $\erf (\cdot)$ denotes the error function. If a forward reaction occurs, the overlapped $A$ molecule is removed and a new $C$ molecule is placed on the surface of the $i$th receptor at the position where the $i$th receptor surface intersects with a straight line describing the displacement of the $A$ molecule, i.e., the line between the positions of the molecule at the beginning and at the end of the simulation step. If $l_2 > \pr \left( \text{Reaction} \, k_f \right)$, then the overlapping $A$ molecule is returned to its previous position, i.e., the position it had at the beginning of the simulation step. 

$4$) If the final position of an $A$ molecule at the end of a simulation step leads to an overlap with the part of the receiver surface not covered by receptors or with a receptor that has been already occupied with another $A$ molecule in a previous simulation step, then the overlapping $A$ molecule is returned to its previous position, i.e., the position it had at the beginning of the simulation step, see double arrows marked with (c) and (d) in Fig. \ref{Fig.SimulationModel}.   

$5$) For each $C$ molecule, a uniformly distributed random number $l_3 \in [0,1]$ is generated. Then, the backward reaction in (\ref{Eq.RevReaction}) occurs if $l_3 \leq \pr \left(\text{Reaction} \, k_b \right)$, where $\pr \left(\text{Reaction} \, k_b \right)$ is the probability that a given $C$ molecule on the surface of the $i$th receptor reverts back and produces an $A$ molecule outside the receiver. $\pr \left(\text{Reaction} \, k_b \right)$ can be evaluated via (\ref{Eq. ProbUniMolReac}) after substituting $k_d$ with $k_b$. The radial position of the new $A$ molecule, $r_{sample}$, is sampled from the normalized distribution $\Pr \left( \text{Ovr} | \vec{r}, \Delta t \right)r^2 / \rho $ and its angular coordinates, i.e., $\theta$ and $\phi$, are chosen as $\theta = \theta_i$ and $\phi = \phi_i$, respectively, where $\theta_i$ and $\phi_i$ are the angular coordinates of the intersection point of the corresponding $A$ molecule with the $i$th receptor, see the single arrow marked with (b) in Fig. \ref{Fig.SimulationModel}. 

\begin{remark} 
In order to be consistent with our assumption of circular receptors, for evaluation of the analytical expressions derived in the previous sections, we consider an \emph{equivalent} circular patch whose area, $S_{\textrm{cir}}^{\textrm{eq}}$, is the same as the area of a triangle receptor used for simulation, $S_{\textrm{tri}}$. Then, the equivalent radius of the circular receptors is given by $r_s^{\textrm{eq}} = \sqrt{S_{\textrm{tri}} / \pi}$. We show the accuracy of this approximation in the following section.  
\end{remark}  
\section{simulation results}
\label{Sec. Simulations}
In this subsection, we present simulation and analytical results for evaluation of the accuracy of the derived closed-form expressions for the channel impulse response. We performed the simulations under two different assumptions. In particular, for the simulation results in Subsections \ref{SubSec. ImpReactionRates} and \ref{SubSec. ImpFiniteRec}, we did not take into account the receptor occupancy in the proposed simulation algorithm. In order to exclude this effect, in step 4) of the simulation framework proposed in Section \ref{Sec. SimulationFramework}, we assumed that any given $A$ molecule can react with a receptor even if it is already occupied. In contrast, for the simulation results in Subsection \ref{SubSec. ImpRecOccupancy}, the effect of receptor occupancy was included. This approach facilitates the assessment of the impact of receptor occupancy on the accuracy of the proposed analytical expressions for the channel impulse response, which does not include the effect of receptor occupancy.  
   
For all simulation results, we adopted $a = 0.5 \, \mu$m and $r_0 = 1 \, \mu$m, i.e., $\di{r}_0 = 2$, and $N_A = 5000$, unless stated otherwise. Furthermore, we assumed that the diffusion coefficient of the information $A$ molecules is $D_A = 5 \times 10^{-9} \, \frac{\text{m}^2}{\text{s}}$. The only parameters that we varied were $\dkd$, $\dkf$, $\dkb$, $M$, and the surface area of the receptors, i.e., $S_{\textrm{cir}}^{\textrm{eq}}$.  

For all results presented, the number of $C$ molecules expected on the surface of the receiver, i.e., $\overline{N}^{\prime}_C(\di{t})$, was evaluated via \eqref{Eq. ExpectedReceSignal}. The simulation results were averaged over $5 \times 10^{4}$ independent releases of $N_A$ $A$ molecules at the transmitter and a simulation step size of $\Delta t = 0.5 \times  10^{-7}$s was chosen.

\subsection{Impact of Reaction Rate Constants} 
\label{SubSec. ImpReactionRates} 
In this subsection, we study the impact of the two \emph{chemical} reaction mechanisms introduced in Section \ref{Sec.SysMod}, i.e., the first order degradation reaction \eqref{Eq.DegReaction} and the second-order reversible reaction \eqref{Eq.RevReaction} on the surface of the receiver, on the received signal at the receiver. We note that the values of $\dkf$, $\dkb$, and $\dkd$ were chosen such that the impact of changing any of these parameters can be observed over the time scale that is simulated. 

\ifdouble
\begin{figure}[!t]  
	\begin{center}
	\includegraphics[scale = 0.33]{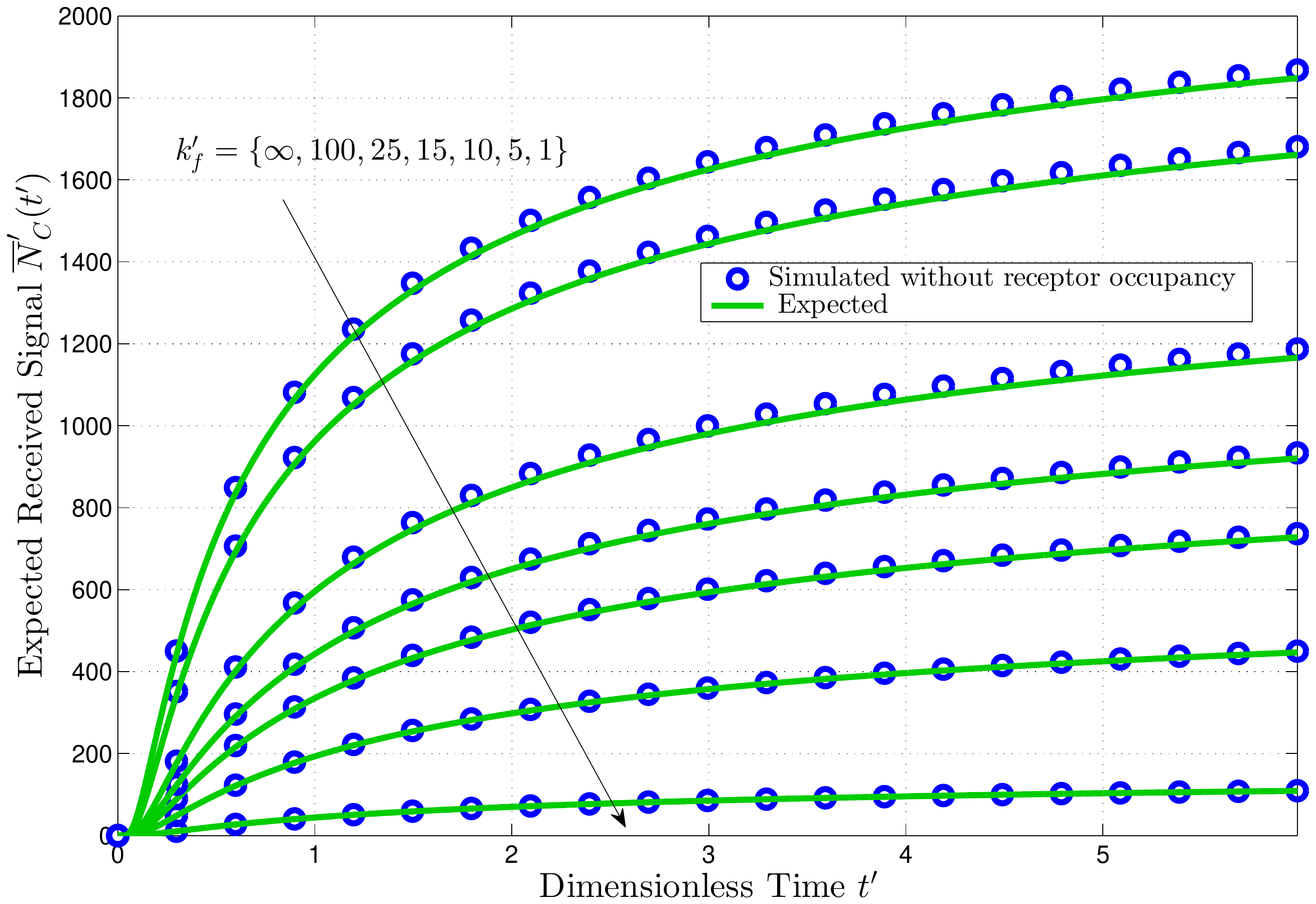}
	\caption{$\overline{N}^{\prime}_C(\di{t})$ as a function of dimensionless time $\di{t}$; impact of $\dkf$ on the expected received signal when $\dkb = \dkd = 0$.} 
	\label{Fig.Forward}
	\end{center}
\end{figure}
\fi  
In Fig. \ref{Fig.Forward}, the number of $C$ molecules expected on the surface of the receiver, $\overline{N}^{\prime}_C(\di{t})$, is shown as a function of dimensionless time $\di{t}$ for dimensionless system parameters $\dkb = \dkd = 0$, and $\dkf = \{1, 5, 10, 15, 25, 100, \infty \}$. Fig. \ref{Fig.Forward} shows that when $\dkb$ is zero, $\overline{N}^{\prime}_C(\di{t})$ is an increasing function in $\di{t}$. This is due to the fact that when $\dkb = 0$ backward reactions do not occur and as soon as an $A$ molecule reacts with a receptor $B$ molecule, the produced $C$ molecule remains on the surface of the receiver and never reverses back to produce a new $A$ molecule. Furthermore, we can observe the impact of decreasing $\dkf$ on the expected received signal. Clearly, when $\dkf$ decreases the probability of a forward reaction happening decreases as well. As a result of this, a given $A$ molecule in the vicinity of the receiver may have multiple contacts with $B$ molecules without a successful reaction happening and finally escape from the proximity of the receiver surface. We note the excellent match between analytical and simulation results. 

\ifdouble
\begin{figure}[!t]
	\vspace*{-1 mm} 
	\begin{center}
	\includegraphics[scale = 0.33]{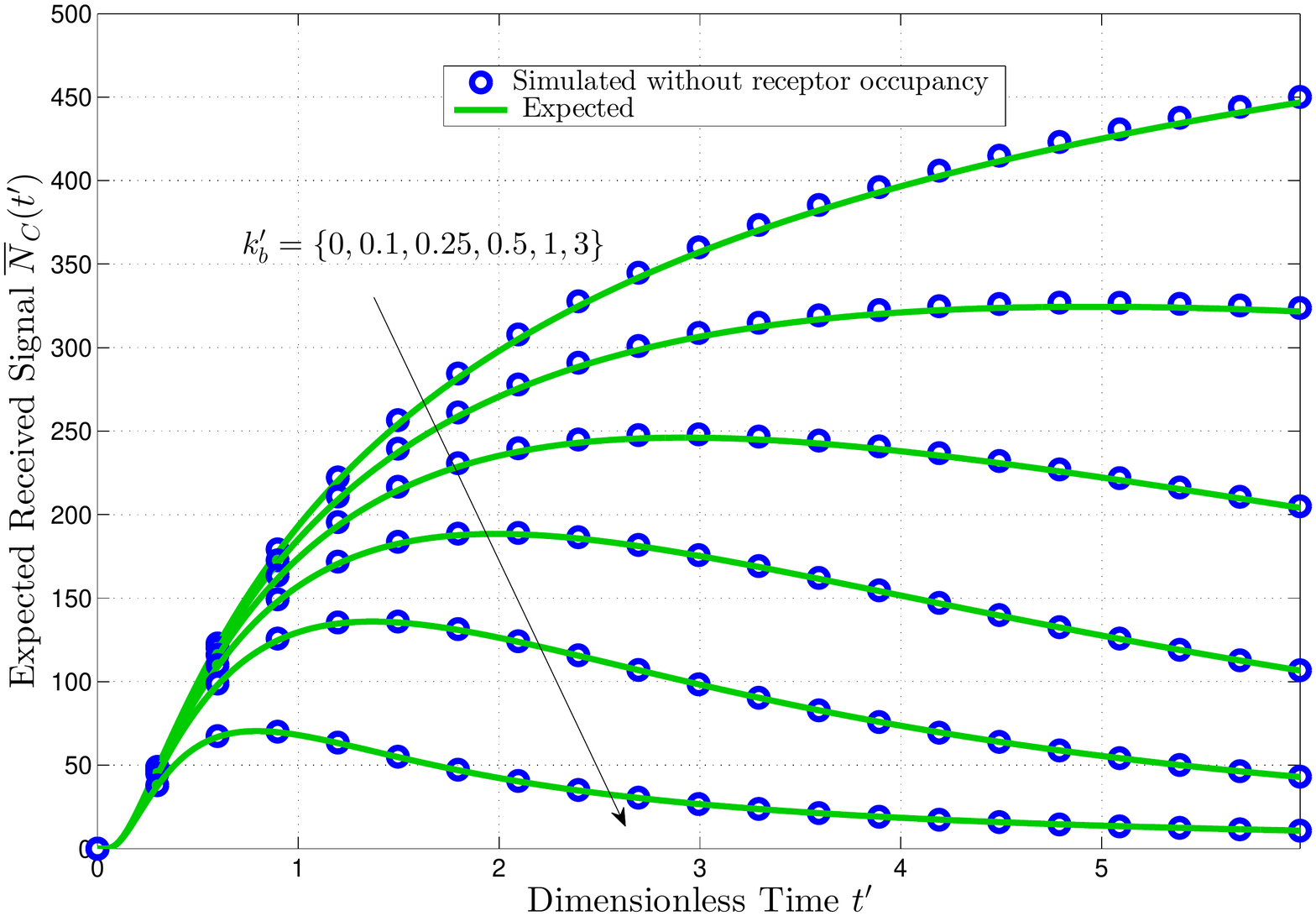}
	\caption{$\overline{N}^{\prime}_C(\di{t})$ as a function of dimensionless time $\di{t}$; impact of $\dkb$ on the expected received signal when $\dkd =0$.} 
	\label{Fig.ForwardBackward}
	\end{center}
\end{figure}
\fi  
In Fig. \ref{Fig.ForwardBackward}, the number of $C$ molecules expected on the surface of the receiver, $\overline{N}^{\prime}_C(\di{t})$, is shown as a function of dimensionless time $\di{t}$ for dimensionless system parameters $\dkd = 0$, $\dkf = 5$, and $\dkb = \{0, 0.1, 0.25, 0.5, 1, 3 \}$. Fig. \ref{Fig.ForwardBackward} reveals that when $\dkb > 0$, $\overline{N}^{\prime}_C(\di{t})$ eventually decreases with increasing $\di{t}$, since, in this case, for any $C$ molecule, there is a non-zero probability that it may reverse back and produce an $A$ molecule in the vicinity of the receiver surface. This new $A$ molecule may associate again with a $B$ molecule on the boundary of the receiver to produce a new $C$ molecule, or it may diffuse away from the receiver and not contribute to the production of another $C$ molecule. We can also observe that the received signal decreases sooner and at a faster rate for larger $\dkb$. This is because increasing $\dkb$ increases the rate at which the $C$ molecules revert back to $A$ molecules (produced by the backward reaction in \eqref{Eq.RevReaction}). We note again the excellent match between simulation and analytical results.

\ifdouble
\begin{figure}[!t]  
	\begin{center}
	\includegraphics[scale = 0.33]{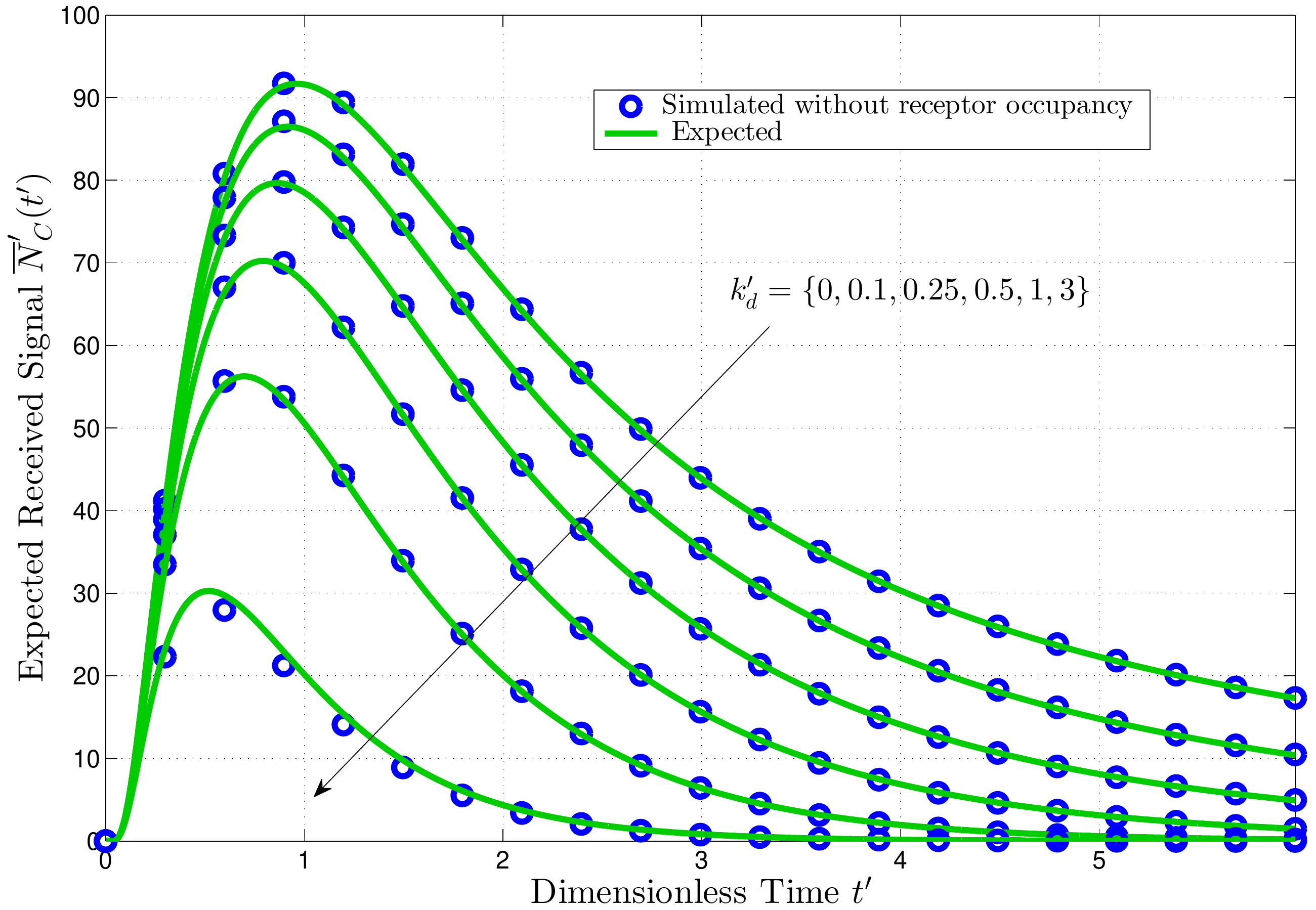}
	\caption{$\overline{N}^{\prime}_C(\di{t})$ as a function of dimensionless time $\di{t}$; impact of $\dkd$.} 
	\label{Fig.ForwardBackwardDeg}
	\end{center}
\end{figure}
\fi  
In Fig. \ref{Fig.ForwardBackwardDeg}, $\overline{N}^{\prime}_C(\di{t})$ is evaluated as a function of dimensionless time $\di{t}$ for dimensionless system parameters $\dkf = 5$, $\dkb = 2$, and $\dkd = \{0, 0.1, 0.25, 0.5, 1, 3 \}$. In this figure, we keep the chemical parameters of the reversible reaction mechanism at the receiver constant to focus on the impact of the degradation reaction in the channel. As expected, $\overline{N}^{\prime}_C(\di{t})$ decreases for increasing $\dkd$. This is because larger values of $\dkd$ increase the probability that a given $A$ molecule degrades in the channel without producing a $C$ molecule at the receiver. As a result, fewer $A$ molecules contribute to the association reaction. 

\subsection{Impact of a Finite Number of Receptors}
\label{SubSec. ImpFiniteRec}
In this subsection, we study the impact on the expected received signal if the fact that the number of receptors in a real system is finite is taken into account. Thereby, the objective is to assess the accuracy of the analytical expressions derived and the approximations introduced in Section \ref{Sec. FiniteRecNum}. For the analysis in this subsection, we consider the case where the total number of receptors that fully covers the surface of the receiver is $M_{\textrm{Max}} = 5120$. The equivalent dimensionless radius of each receptor is $r^{\textrm{eq} \, \prime}_{\textrm{cir}} = 0.0279$.

\ifdouble
\begin{figure}[!t]  
	\begin{center}
	\includegraphics[scale = 0.33]{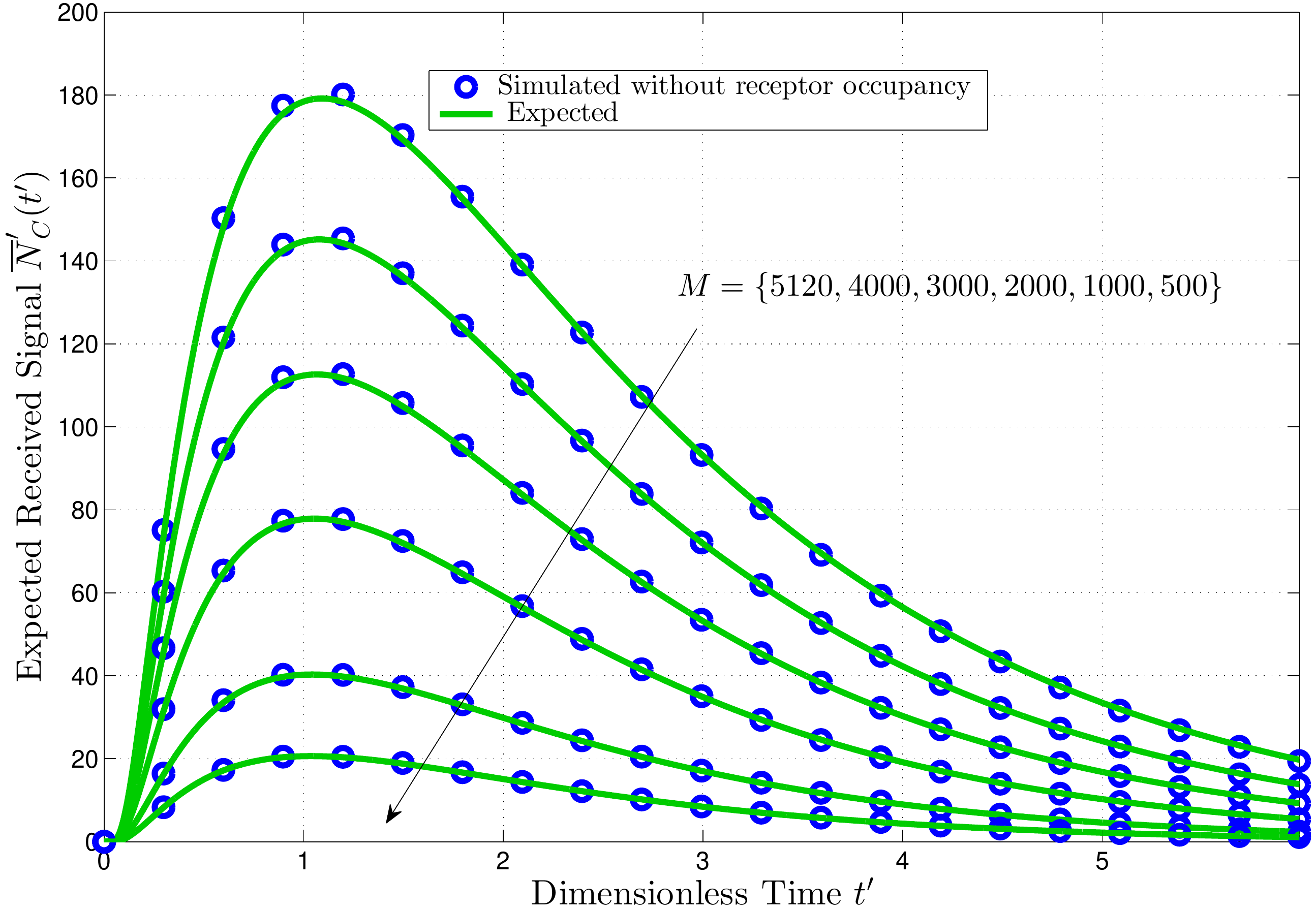}
	\caption{$\overline{N}^{\prime}_C(\di{t})$ as a function of dimensionless time $\di{t}$; impact of the number of receptors, $M$.} 
	\label{Fig.ReceptorAna1}
	\end{center}
\end{figure}
\fi     
In Fig. \ref{Fig.ReceptorAna1}, $\overline{N}^{\prime}_C(\di{t})$ is evaluated as a function of dimensionless time $\di{t}$ for dimensionless system parameters $\dkf = 10$, $\dkb = 1$, and $\dkd = 0.5$ and $M = \{5120, 4000, 3000, 2000, 1000, 500 \}$. As can be seen from Fig. \ref{Fig.ReceptorAna1}, as expected, reducing the number of receptors reduces the overall expected received signal, i.e., $\overline{N}^{\prime}_C(\di{t})$. This is because by decreasing $M$, the area of the receiver that is not covered by receptors increases, and, as a result of this, the probability that a given $A$ molecule close to the surface of the receiver is reflected back and diffuses away increases. Furthermore, when $M$ is large, even if an $A$ molecule is reflected back by the part of the receiver that is not covered by receptors, the probability that the molecule is captured by neighbouring receptors at a future time increases. Fig. \ref{Fig.ReceptorAna1} also shows an excellent match between simulation and analytical results, which confirms the accuracy of the assumptions and approximations introduced in Section \ref{Sec. FiniteRecNum}.                 

\ifdouble
\begin{figure}[!t]
	\vspace*{-0.5 mm} 
	\begin{center}
	\includegraphics[scale = 0.33]{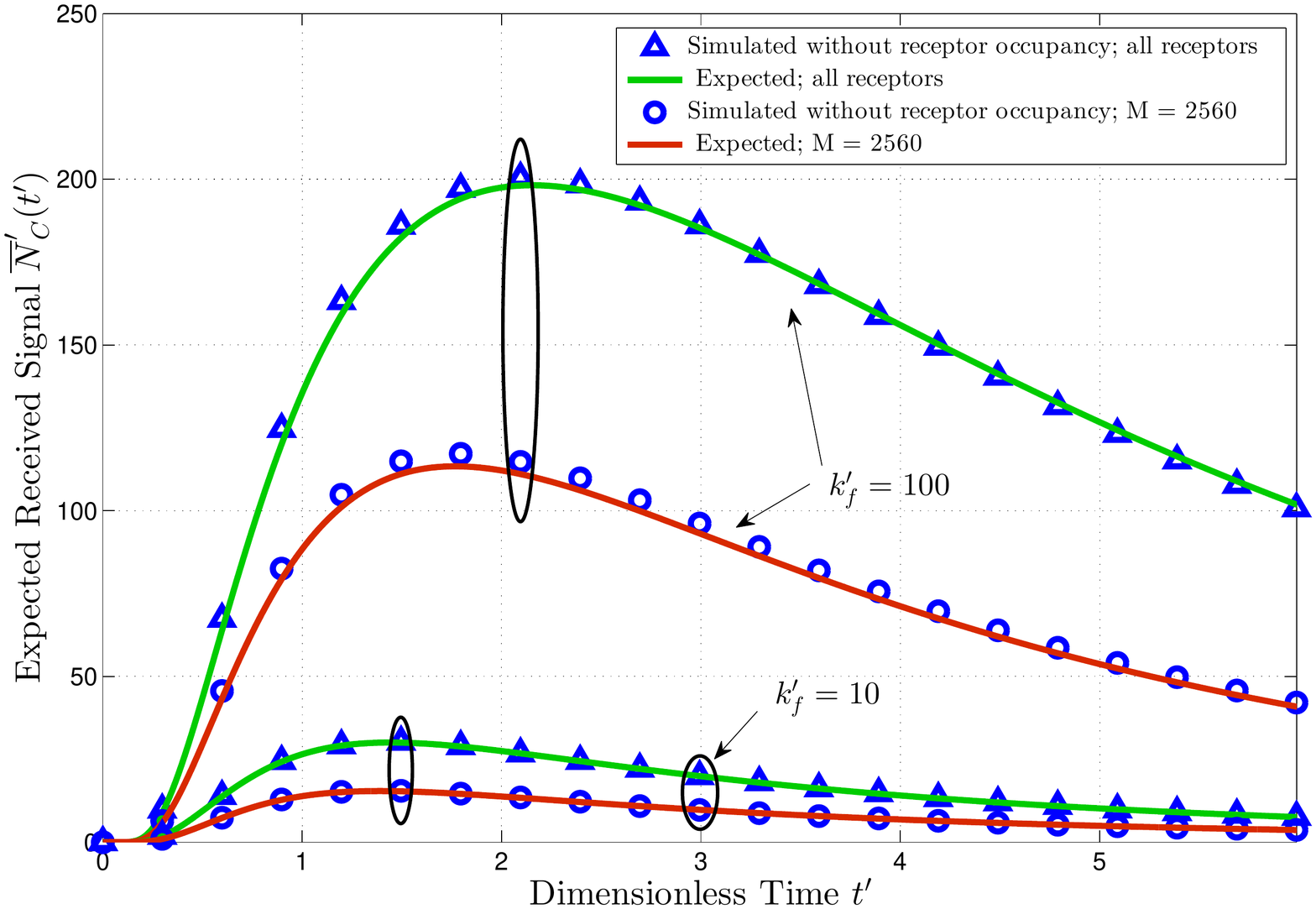}
	\caption{$\overline{N}^{\prime}_C(\di{t})$ as a function of dimensionless time $\di{t}$; impact of forward reaction rate constant, $\dkf$.} 
	\label{Fig.ReceptorAna2}
	\end{center}
\end{figure}
\fi  
In Fig. \ref{Fig.ReceptorAna2}, $\overline{N}^{\prime}_C(\di{t})$ is evaluated as a function of dimensionless time $\di{t}$ for dimensionless system parameters $\dkb = 4$, $\dkd = 0.1$, and $\di{r}_0 = 3$ and $M = \{5120, 2560 \}$. In this figure, we show the impact of $\dkf$ on $\overline{N}^{\prime}_C(\di{t})$ when the number of receptors, $M$, is reduced by a factor of $1/2$ for two different values of $\dkf = \{100, 10 \}$. As already observed from Fig. \ref{Fig.ReceptorAna1}, reducing $M$ reduces $\overline{N}^{\prime}_C(\di{t})$. However, Fig. \ref{Fig.ReceptorAna2}, indicates that this gap, i.e., the gap between $\overline{N}^{\prime}_C(\di{t})$ when all possible receptors are used and $\overline{N}^{\prime}_C(\di{t})$ when only half of the possible receptors are deployed, increases with increasing $\dkf$. This is because when $\dkf$ is large, the probability of forward reaction, and, as a result of this, the overall probability that a given $A$ molecule that hits the surface of the receiver reacts with a receptor $B$ molecule is large. Thus, in this case, reducing $M$ has a more profound impact on the reduction of the overall reaction probability compared to the case when $\dkf$ is small. 

\subsection{Impact of Receptor Occupancy}
\label{SubSec. ImpRecOccupancy}
In this subsection, we study the impact of the effect of receptor occupancy on the expected received signal. We assume that $M_{\textrm{Max}} = 1280$ and $r^{\textrm{eq} \, \prime}_{\textrm{cir}} = 0.0279$, which corresponds to a dimensionless receptor surface area of $\pi \times {r^{\textrm{eq} \, \prime}_{\textrm{cir}}}^{2}$ and we denote this surface area by $S_0$. Furthermore, in order to focus on the impact of receptor occupancy, we assume that $\dkd = 0$. However, similar observations can be made for $\dkd > 0$.          

\ifdouble
\begin{figure}[!t]
	\vspace*{-0.5 mm} 
	\begin{center}
	\includegraphics[scale = 0.33]{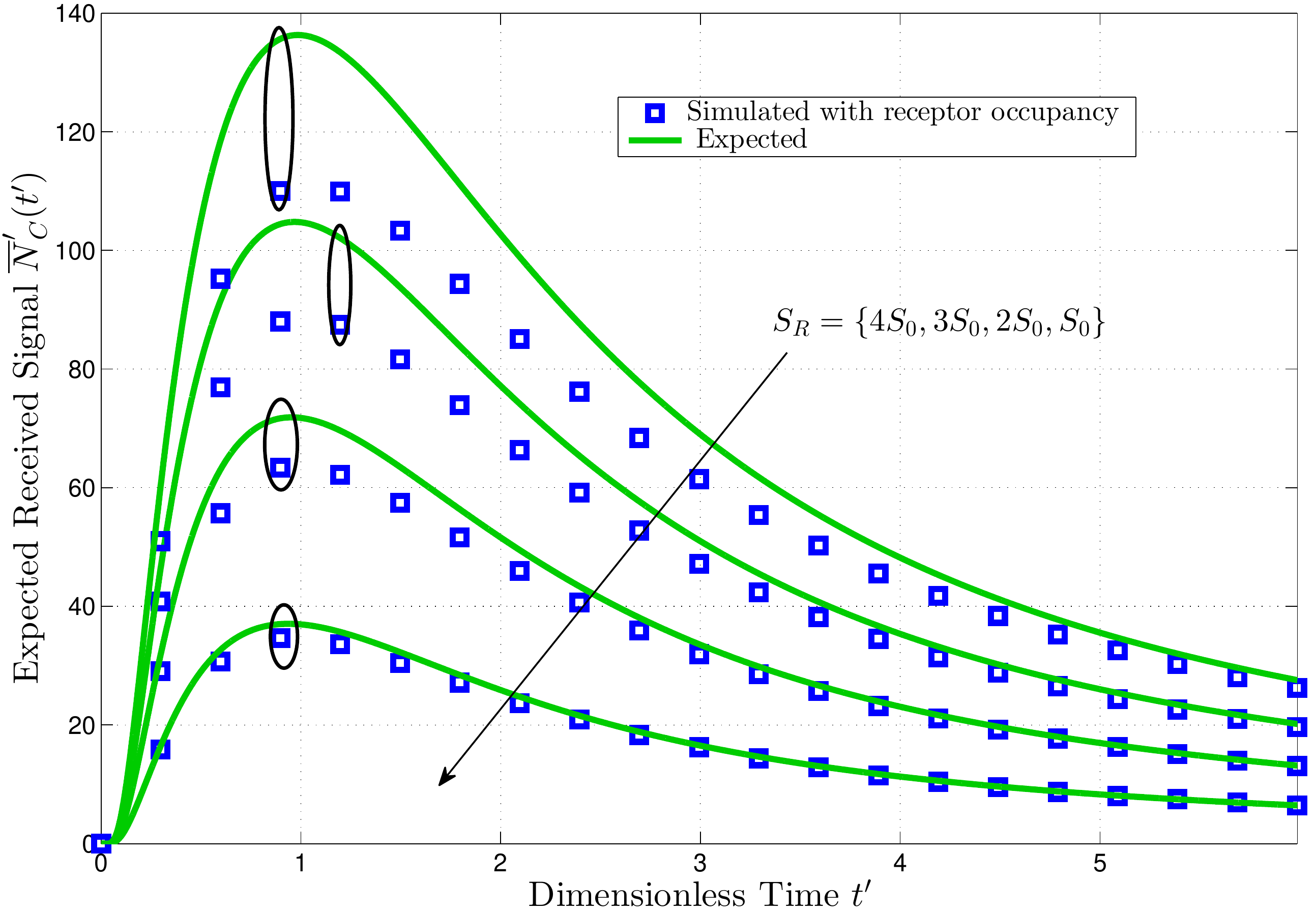}
	\caption{$\overline{N}^{\prime}_C(\di{t})$ as a function of dimensionless time $\di{t}$; impact of receptor surface area, $S_R$.} 
	\label{Fig.RecArea}
	\end{center}
\end{figure}
\fi   
In Fig. \ref{Fig.RecArea}, we show $\overline{N}^{\prime}_C(\di{t})$ versus dimensionless time $\di{t}$ for dimensionless system parameters $\dkb = 2$ and $\dkf = 10$ and $M = 1000$. Fig. \ref{Fig.RecArea} shows that the effect of receptor occupancy causes a deviation between analytical and simulation results. As can be observed, the deviation is larger if the surface areas of individual receptors, $S_R$, is larger, e.g., $S_R = 4S_0$. In particular, the analytical results overestimate the expected number of $C$ molecules. This is due to the fact that when $S_R$ is larger, during the time that a given $A$ molecule is bound to a certain receptor $B$ molecule, the probability that other $A$ molecules, that are in the vicinity of this receptor, hit the occupied receptor and are reflected back increases. Furthermore, we observe that for larger values of $S_R$, the deviation between analytical and simulation results is larger around the value of $\di{t}$ that corresponds to the peak value of $\overline{N}^{\prime}_C(\di{t})$. This is because at this dimensionless time the number of $A$ molecules in the proximity of the receiver is larger and it is more probable that some of them collide with an occupied receptor. On the other hand, when $S_R$ is small, e.g., $S_R = S_0$, the proposed analytical expression provides an excellent approximation for the corresponding simulation result. This suggest that when $S_R$ is small the impact of receptor occupancy is negligible.

\ifdouble
\begin{figure}[!t]
	\vspace*{-0.5 mm} 
	\begin{center}
	\includegraphics[scale = 0.33]{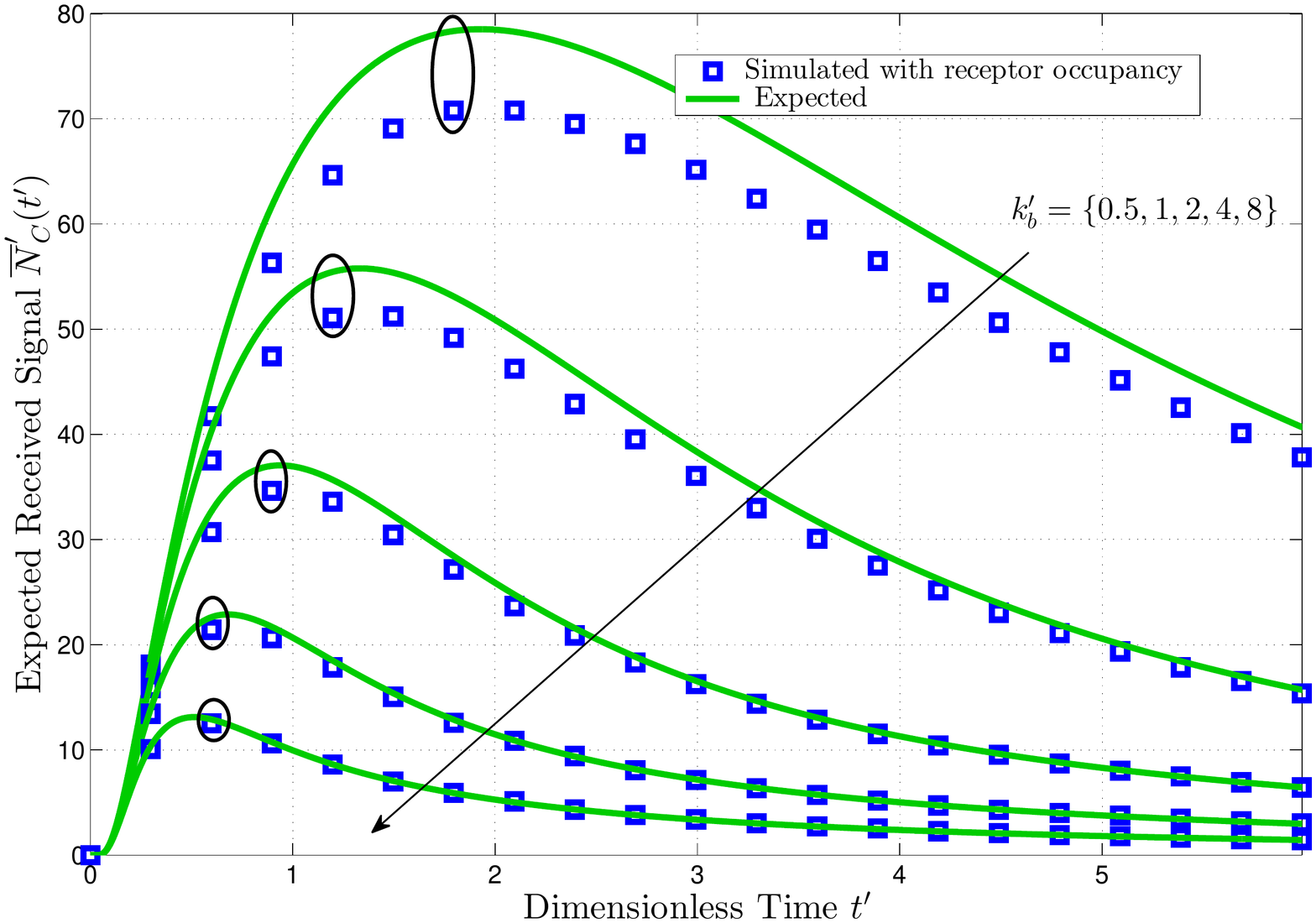}
	\caption{$\overline{N}^{\prime}_C(\di{t})$ as a function of dimensionless time $\di{t}$; impact of dimensionless backward reaction rate constant, $\dkb$.} 
	\label{Fig.RecKb}
	\end{center}
\end{figure}
\fi   
In Fig. \ref{Fig.RecKb}, we depict $\overline{N}^{\prime}_C(\di{t})$ versus dimensionless time $\di{t}$ for dimensionless system parameters $\dkf = 10$, $\dkb = \{0.5, 1, 2, 4, 8 \}$, and $S_R = S_0$ and $M = 1000$. As can be observed, increasing $\dkb$ reduces the deviation between analytical and simulation results caused by the impact of receptor occupancy. This is due to the fact that for larger values of $\dkb$, the mean occupancy time for an $A$ molecule bound to a receptor $B$ molecule decreases. As a result, it is less likely that $A$ molecules in the vicinity of the receiver collide with an occupied receptor. This also shows that the proposed analytical expression provides an accurate approximation for larger values of $\dkb$. 

\ifdouble
\begin{figure}[!t] 
	\vspace*{-0.5 mm} 
	\centering
	\includegraphics[scale = 0.33]{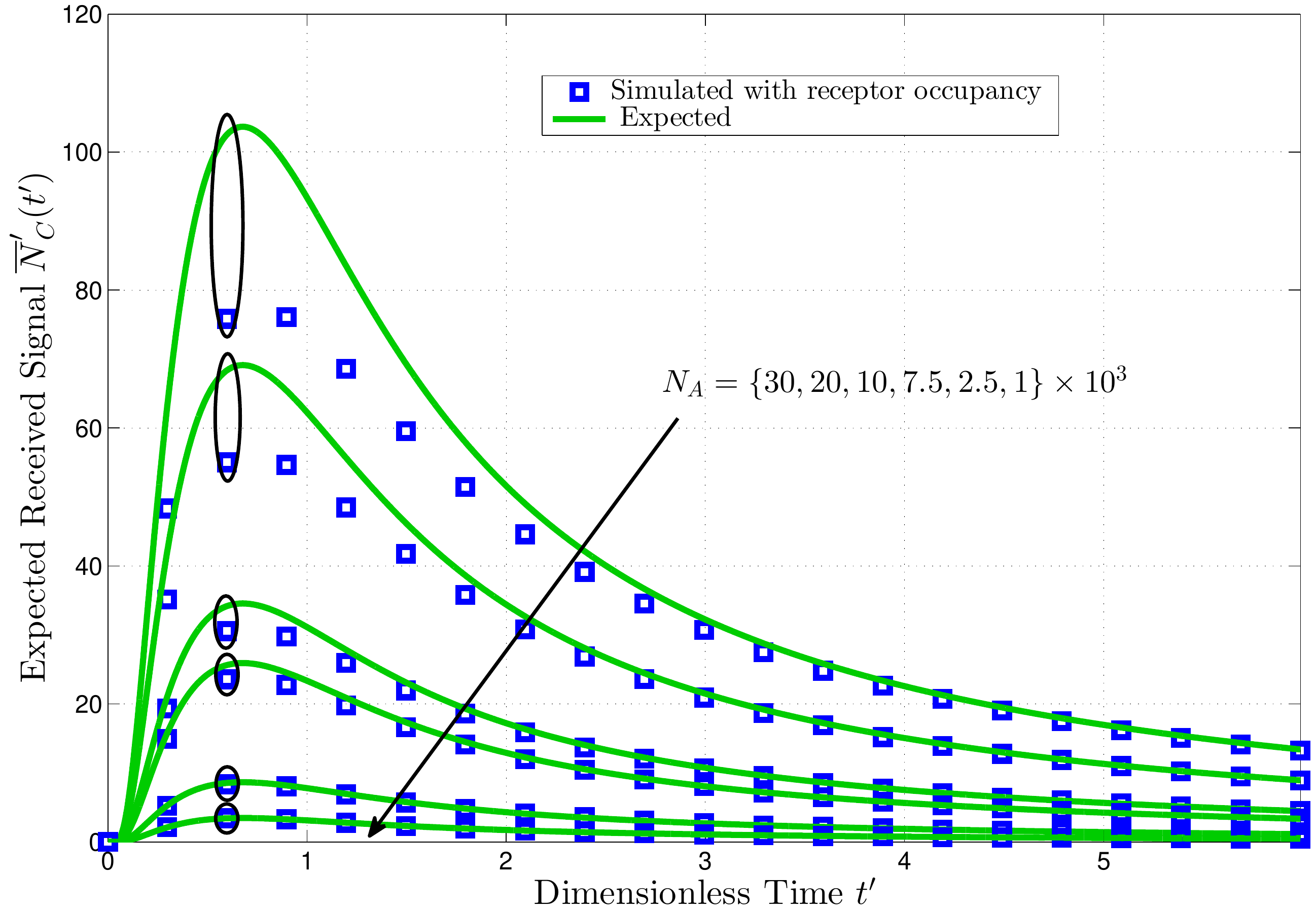}
	\caption{$\overline{N}^{\prime}_C(\di{t})$ as a function of dimensionless time $\di{t}$; impact of the number of released molecules, $N_A$.}
	\label{Fig.RecNa}
\end{figure}
\else
\begin{figure}[!t]
	\centering
	\includegraphics[scale = 0.34]{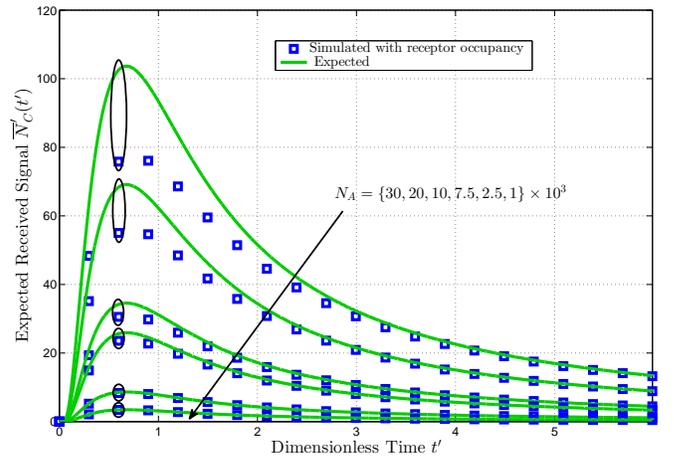}\vspace*{-4 mm}
	\caption{$\overline{N}^{\prime}_C(\di{t})$ as a function of dimensionless time $\di{t}$; impact of the number of released molecules, $N_A$.}\vspace*{-5 mm} 
	\label{Fig.RecNa}
\end{figure}
\fi 
In Fig. \ref{Fig.RecNa}, $\overline{N}^{\prime}_C(\di{t})$ is evaluated as a function of dimensionless time $\di{t}$ for dimensionless system parameters $\dkf = 10$, $\dkb = 4$, and $S_R = S_0$. We furthermore assumed that $M = 750$ and $N_A = \{1000, 2500, 7500, 10000, 20000, 30000 \}$. Fig. \ref{Fig.RecNa} shows that for increasing $N_A$, the deviation between the analytical and simulation results increases. This is because by releasing more $A$ molecules at the transmitter side, more $A$ molecules arrive in the vicinity of the receiver. As a result of this, the probability that some of these $A$ molecules collide with an occupied receptor increases.                  
\section{conclusions} 
In this paper, we considered a diffusive molecular communication channel between a pair of transmitter and receiver nano-machines. We modelled the reception at the receiver as a second-order reaction mechanism, where the information molecules released by the transmitter into the fluid environment could reversibly react with the receptor protein molecules covering the surface of the receiver. We considered two different scenarios. In the first scenario, we assumed that receptor protein molecules covered the entire surface of the receiver, whereas, in the second scenario, they partially covered the receiver surface. We furthermore assumed that the information molecules may degrade in the channel via a first-order degradation reaction. We derived a closed-form \emph{time domain} expression for the channel impulse response of the system for both considered scenarios and verified its accuracy via particle-based simulations. An excellent match between analytical and simulation results was observed for both considered scenarios when the effect of receptor occupancy was negligible. The derived channel impulse response can be employed as a basic building block for design of functionalities such as modulation, detection, and estimation in diffusive MC systems.    

In our future work, we plan to expand our analysis to include additional characteristics of biological receptors such as the impact of receptor clustering, receptor internalization, and receptor-drug interactions.
\label{Sec.Con}  
\appendix[Proof of Theorem 1]
We denote the Fourier, inverse Fourier, Laplace, and inverse Laplace transforms by $F\{\cdot\}$, $F^{-1}\{\cdot\}$, $L\{\cdot\}$, and $L^{-1}\{\cdot\}$, respectively.

We start by solving reaction-diffusion equation (\ref{Eq. Reaction-DiffusionU}) for initial condition (\ref{Eq. InitialConditionU}). Taking the Fourier transform of (\ref{Eq. Reaction-DiffusionU}) with respect to $\di{r}$ leads to the following partial differential equation 
\begin{IEEEeqnarray}{c}
	\label{Eq. AppFourierPartial} 
	\frac{\partial \overline{U}(\kappa, \di{t} | \di{r}_0)}{\partial \di{t}} = -(\kappa^2 + \dkd ) \overline{U}(\kappa,\di{t} | \di{r}_0). 
 \end{IEEEeqnarray} 
where $\overline{U}(\kappa,\di{t} | \di{r}_0) = F\{\di{r}\di{U}(\di{r},\di{t}|\di{r}_0)\}$, with $\kappa$ representing the Fourier domain variable, is given by 
\begin{IEEEeqnarray}{c} 
	\label{Eq. AppFourierTransform}
	\hspace{-4 mm} \overline{U}(\kappa,\di{t} | \di{r}_0) = \frac{1}{2\pi} \int_{- \infty}^{+ \infty} \di{r}\di{U}(\di{r},\di{t}|\di{r}_0) \exp(-j\kappa \di{r}) \dif \di{r}. 
\end{IEEEeqnarray}
 
Solving \eqref{Eq. AppFourierPartial} for $\overline{U}(\kappa,\di{t} | \di{r}_0)$ leads to 
\begin{IEEEeqnarray}{c} 
	\label{Eq. App1} 
	\overline{U}(\kappa,\di{t} | \di{r}_0) = C_0 \exp( -\kappa^2 \di{t} + \dkd \di{t}),
\end{IEEEeqnarray} 
where $C_0$ is a constant to be determined by the initial condition. Taking the Fourier transform of \eqref{Eq. InitialConditionU} and employing (\ref{Eq. App1}), constant $C_0$ is obtained as 
\begin{IEEEeqnarray}{c}
	\label{Eq. AppConstantC} 
	C_0 = \frac{\exp(-j\kappa \di{r}_0)}{4 \pi \di{r}_0}.
\end{IEEEeqnarray}

Finally, $\di{r}\di{U}(\di{r},\di{t}|\di{r}_0)$ can be evaluated as 
\ifdouble 
\begin{IEEEeqnarray}{rCl}
	\label{Eq. AppU_time} 
	\di{r}\di{U}(\di{r},\di{t}|\di{r}_0) & = & F^{-1}\{\overline{U}(\kappa,\di{t} | \di{r}_0)\} \nonumber \\
	& = & \frac{\exp(-\dkd \di{t})}{8 \pi \di{r}_0 \sqrt{\pi \di{t}}} \exp \left( \frac{-(\di{r} - \di{r}_0)^2}{4 \di{t}} \right).  
\end{IEEEeqnarray}
\else
\begin{IEEEeqnarray}{C}
	\label{Eq. AppU_time} 
	\di{r}\di{U}(\di{r},\di{t}|\di{r}_0) = F^{-1}\{\overline{U}(\kappa,\di{t} | \di{r}_0)\} 
	 = \frac{\exp(-\dkd \di{t})}{8 \pi \di{r}_0 \sqrt{\pi \di{t}}} \exp \left( \frac{-(\di{r} - \di{r}_0)^2}{4 \di{t}} \right).
\end{IEEEeqnarray}  
\fi 

Next, we solve reaction-diffusion equation (\ref{Eq. Reaction-DiffusionV}) for initial condition (\ref{Eq. ConditionV}). In order to do so, we first apply the Laplace transform with respect to $\di{t}$ to (\ref{Eq. Reaction-DiffusionV}) which results in
\ifdouble 
\begin{IEEEeqnarray}{rCl}
	\label{Eq. App2} 
	s\di{r}\overline{V}(\di{r},s|\di{r}_0) - \di{r}\di{V}(\di{r},\di{t} = 0 |\di{r}_0) & = &  \frac{\partial^2}{\partial r^{\prime^{2}}} \big(\di{r} \overline{V}(\di{r},s|\di{r}_0) \big) \nonumber \\ 
	&& -\> \dkd \di{r} \overline{V}(\di{r},s|\di{r}_0), \nonumber \\*
\end{IEEEeqnarray}
\else
\begin{IEEEeqnarray}{rCl}
	\label{Eq. App2} 
	s\di{r}\overline{V}(\di{r},s|\di{r}_0) - \di{r}\di{V}(\di{r},\di{t} = 0 |\di{r}_0) & = &  \frac{\partial^2}{\partial r^{\prime^{2}}} \big(\di{r} \overline{V}(\di{r},s|\di{r}_0) \big) 
	 -\dkd \di{r} \overline{V}(\di{r},s|\di{r}_0),
\end{IEEEeqnarray}
\fi
where $\overline{V}(\di{r},s|\di{r}_0) = L\{\di{V}(\di{r},\di{t}|\di{r}_0)\}$, i.e., 
\begin{IEEEeqnarray}{c}
	\label{Eq. AppLaplaceTransform} 
	\overline{V}(\di{r},s|\di{r}_0) = \int_{0}^{\infty} \di{V}(\di{r},\di{t}|\di{r}_0) \exp(-s\di{t}) \dif \di{t}. 
\end{IEEEeqnarray}

The second term on the left hand side of (\ref{Eq. App2}) is zero (see (\ref{Eq. ConditionV})). Eq. (\ref{Eq. App2}) is a partial differential equation of function $\di{r}\overline{V}(\di{r},s|\di{r}_0)$. As can be seen from \eqref{Eq. AppU_time}, $\di{U}(\di{r} \to \infty,\di{t} | \di{r}_0) =0$. Thus, in order to satisfy (\ref{Eq. BounCon1}), $\di{V}(\di{r} \to \infty, \di{t} | \di{r}_0) = 0$. Taking this into account, the solution of \eqref{Eq. App2} can be written as 
\begin{IEEEeqnarray}{c}
	\label{Eq. AppV_L} 
	\overline{V}(\di{r},s|\di{r}_0) = \frac{q}{\di{r}} \exp\left( - \di{r} \sqrt{s + \dkd} \right),
\end{IEEEeqnarray}
where $q$ is a constant that can be used to ensure that $\di{U}(\di{r},\di{t} | \di{r}_0)$ and $\di{V}(\di{r},\di{t} | \di{r}_0)$ jointly satisfy boundary condition (\ref{Eq. DiBounCon2}). In order to find $q$, we first evaluate $L\{\di{P}_A(\di{r},\di{t} | \di{r}_0)\} = \overline{P}_A(\di{r},s | \di{r}_0) = L\{\di{U}(\di{r},\di{t} | \di{r}_0)\} + L\{\di{V}(\di{r},\di{t}| \di{r}_0)\}$ which yields
\ifdouble
\begin{IEEEeqnarray}{rCl} 
 \label{Eq. AppCompleteSolution_L} 
 \overline{P}_A(\di{r},s | \di{r}_0) & = & \frac{\exp\left( - \sqrt{s + \dkd} ( \di{r} - \di{r}_0 ) \right)}{8 \pi \di{r} \di{r}_0 \sqrt{s + \dkd}} \nonumber \\
 && +\> \frac{q}{\di{r}} \exp\left( - \di{r} \sqrt{s + \dkd} \right),
\end{IEEEeqnarray}
\else
\begin{IEEEeqnarray}{rCl} 
 \label{Eq. AppCompleteSolution_L} 
 \overline{P}_A(\di{r},s | \di{r}_0) & = & \frac{\exp\left( - \sqrt{s + \dkd} ( \di{r} - \di{r}_0 ) \right)}{8 \pi \di{r} \di{r}_0 \sqrt{s + \dkd}} 
  + \frac{q}{\di{r}} \exp\left( - \di{r} \sqrt{s + \dkd} \right),
\end{IEEEeqnarray}
\fi
where we used \cite[Eq. 29.3.84]{abramowitz} 
\begin{IEEEeqnarray}{c} 
	L \left\lbrace \frac{1}{\sqrt{\pi \di{t}}} \exp\left( \frac{-b^2}{4\di{t}} \right) \right\rbrace = \frac{\exp(-b \sqrt{s})}{\sqrt{s}} 
\end{IEEEeqnarray}
for evaluation of $L\{\di{U}(\di{r},\di{t} | \di{r}_0)\}$. Then, we calculate the Laplace transform of boundary condition (\ref{Eq. DiBounCon2}), which leads to 
\begin{IEEEeqnarray}{c} 
	\label{Eq. AppBoundaryCon2_L} 
	\frac{\partial \overline{P}_A(\di{r},s | \di{r}_0)}{\partial \di{r}} \bigg|_{\di{r} = 1} = \frac{\dkf s}{ 4 \pi s + \dkb 4 \pi} \overline{P}_A(\di{r},s | \di{r}_0). 
\end{IEEEeqnarray}

Now, taking the derivative of (\ref{Eq. AppCompleteSolution_L}) with respect to $\di{r}$, substituting the resulting equation into (\ref{Eq. AppBoundaryCon2_L}), and solving the corresponding equation for $q$ yields
\ifdouble 
\begin{IEEEeqnarray}{rCl} 
	\label{Eq. AppConstantq} 
	q & = & \frac{ 4 \pi \sqrt{s+\dkd} - 4 \pi - \frac{\dkf s}{s + \dkb}}{4 \pi \sqrt{s+\dkd} + 4 \pi + \frac{\dkf s}{s + \dkb}} \times \frac{1}{8 \pi \di{r}_0 \sqrt{s + \dkd}} \nonumber \\
	&& \times \> \exp\left( - \sqrt{s+\dkd} (\di{r}_0 - 2) \right).    
\end{IEEEeqnarray} 
\else
\begin{IEEEeqnarray}{rCl} 
	\label{Eq. AppConstantq} 
	q & = & \frac{ 4 \pi \sqrt{s+\dkd} - 4 \pi - \frac{\dkf s}{s + \dkb}}{4 \pi \sqrt{s+\dkd} + 4 \pi + \frac{\dkf s}{s + \dkb}} \times \frac{1}{8 \pi \di{r}_0 \sqrt{s + \dkd}}
	 \exp\left( - \sqrt{s+\dkd} (\di{r}_0 - 2) \right).    
\end{IEEEeqnarray} 
\fi

The Laplace transform of the final solution can be evaluated by substituting (\ref{Eq. AppConstantq}) into (\ref{Eq. AppCompleteSolution_L}), and can be written as the sum of three terms as follows:
\ifdouble
\begin{IEEEeqnarray}{lCl}  
	\label{Eq. AppCompleteSolution_L_and_q} 
	 \overline{P}_A(\di{r},s | \di{r}_0)  & = & \frac{\exp\left( - \sqrt{s + \dkd} ( \di{r} - \di{r}_0 ) \right)}{8 \pi \di{r} \di{r}_0 \sqrt{s + \dkd}} \nonumber \\ 
	&& +\> \frac{\exp\left( - \sqrt{s + \dkd} ( \di{r} + \di{r}_0 -2 ) \right)}{8 \pi \di{r} \di{r}_0 \sqrt{s + \dkd}} \nonumber \\  
	&& -\> \left[ \frac{4 \pi + \frac{s \dkf}{s + \dkb}}{4 \pi \sqrt{s + \dkd} + 4 \pi + \frac{s \dkf}{ s + \dkb}} \right. \nonumber \\ 
	&& \left. \times\> \frac{\exp\left( - \sqrt{s + \dkd} ( \di{r} + \di{r}_0 -2 ) \right)}{4 \pi \di{r} \di{r}_0 \sqrt{s + \dkd}} \right].
\end{IEEEeqnarray}
\else
\begin{IEEEeqnarray}{lCl}  
	\label{Eq. AppCompleteSolution_L_and_q} 
	 \overline{P}_A(\di{r},s | \di{r}_0)  & = & \frac{\exp\left( - \sqrt{s + \dkd} ( \di{r} - \di{r}_0 ) \right)}{8 \pi \di{r} \di{r}_0 \sqrt{s + \dkd}}  
	+ \frac{\exp\left( - \sqrt{s + \dkd} ( \di{r} + \di{r}_0 -2 ) \right)}{8 \pi \di{r} \di{r}_0 \sqrt{s + \dkd}} \nonumber \\  
	&& -\> \left[ \frac{4 \pi + \frac{s \dkf}{s + \dkb}}{4 \pi \sqrt{s + \dkd} + 4 \pi + \frac{s \dkf}{ s + \dkb}} 
	 \times\> \frac{\exp\left( - \sqrt{s + \dkd} ( \di{r} + \di{r}_0 -2 ) \right)}{4 \pi \di{r} \di{r}_0 \sqrt{s + \dkd}} \right].
\end{IEEEeqnarray}
\fi

Taking the inverse Laplace transform of the first two terms on the right hand side of (\ref{Eq. AppCompleteSolution_L_and_q}) yields the first two terms on the right hand side of (\ref{Eq. Green's function}). The denominator of the third term on the right hand side in (\ref{Eq. AppCompleteSolution_L_and_q}) can be rearranged and written as 
\ifdouble 
\begin{IEEEeqnarray}{c} 
	\label{Eq. AppDenominator} 
	\left[ (s \hs + \hs \dkd)^{3/2} + (\dkb - \dkd)\sqrt{s \hs + \hs \dkd} - \frac{4 \pi \hs + \hs \dkf}{4 \pi} \dkd \right. \nonumber \\ 
	\left. + \frac{4 \pi \hs + \hs \dkf}{4 \pi}(s \hs + \hs \dkd) + \dkb \right] \left(4 \pi \di{r} \di{r}_0 \sqrt{s \hs + \hs \dkd)} \right).
\end{IEEEeqnarray}
\else
\begin{IEEEeqnarray}{c} 
	\label{Eq. AppDenominator} 
	\left[(s + \dkd)^{3/2} + (\dkb - \dkd)\sqrt{s +\dkd} - \frac{4 \pi + \dkf}{4 \pi} \dkd \right. \nonumber \\ 
	\left. + \frac{4 \pi + \dkf}{4 \pi}(s + \dkd) + \dkb \right] \left(4 \pi \di{r} \di{r}_0 \sqrt{s + \dkd)} \right). 
\end{IEEEeqnarray}
\fi 

The terms in brackets can be interpreted as a cubic equation in $\sqrt{s+\dkd}$. Let us assume that $-\di{\alpha}$, $-\di{\beta}$, and $-\di{\gamma}$ are the roots of this cubic equation. Then, it can be easily verified that these roots have to satisfy the system of equations in (\ref{Eq. AlphaBetaGamma}). Employing the partial fraction expansion technique, the third term on the right hand side of (\ref{Eq. AppCompleteSolution_L_and_q}) can be written as
\ifdouble 
\begin{IEEEeqnarray}{C} 
	\label{Eq. AppLastTermDecompose} 
	\left( \frac{\di{\eta}_1}{\di{\alpha} + \sqrt{s + \dkd}} + \frac{\di{\eta}_2}{\di{\beta} + \sqrt{s + \dkd}} + \frac{\di{\eta}_3}{\di{\gamma} + \sqrt{s + \dkd}} \right)  \nonumber\\ 
	  \times \frac{1}{4 \pi \di{r} \di{r}_0 \sqrt{s + \dkd}} \exp\left( - \sqrt{s + \dkd}( \di{r} + \di{r}_0 -2 ) \right), 
\end{IEEEeqnarray}
\else
\begin{IEEEeqnarray}{C} 
	\label{Eq. AppLastTermDecompose} 
	\left( \frac{\di{\eta}_1}{\di{\alpha} + \sqrt{s + \dkd}} + \frac{\di{\eta}_2}{\di{\beta} + \sqrt{s + \dkd}} + \frac{\di{\eta}_3}{\di{\gamma} + \sqrt{s + \dkd}} \right)  \nonumber\\ 
	  \times \frac{1}{4 \pi \di{r} \di{r}_0 \sqrt{s + \dkd}} \exp\left( - \sqrt{s + \dkd}( \di{r} + \di{r}_0 -2 ) \right),
\end{IEEEeqnarray}
\fi
where $\di{\eta}_1$, $\di{\eta}_2$, and $\di{\eta}_3$ are the residues and given by (\ref{Eq. eta1})-(\ref{Eq. eta3}). Taking the inverse Laplace transform of (\ref{Eq. AppLastTermDecompose}) results in the last three terms on the right hand side of (\ref{Eq. Green's function}), where we used \cite[Eq. 29.3.90]{abramowitz}
\ifdouble 
\begin{IEEEeqnarray}{rCl} 
	\label{Eq. AppLaplaceInverseLastStep} 
	L^{-1}\left\lbrace \frac{\exp(-n\sqrt{s})}{\sqrt{s}(m + \sqrt{s})} \right\rbrace & = & \exp\left( nm + m^2 t \right) \nonumber \\ 
	 && \times\> \erfc\left( \frac{n}{2\sqrt{t}} + m\sqrt{t} \right).
\end{IEEEeqnarray} 
\else
\begin{IEEEeqnarray}{rCl} 
	\label{Eq. AppLaplaceInverseLastStep} 
	L^{-1}\left\lbrace \frac{\exp(-n\sqrt{s})}{\sqrt{s}(m + \sqrt{s})} \right\rbrace & = & \exp\left( nm + m^2 \di{t} \right)  
	 \times \erfc\left( \frac{n}{2\sqrt{\di{t}}} + m\sqrt{t} \right).
\end{IEEEeqnarray} 
\fi         
\bibliography{IEEEabrv,Library}

\begin{thebibliography}{10}
\providecommand{\url}[1]{#1}
\csname url@samestyle\endcsname
\providecommand{\newblock}{\relax}
\providecommand{\bibinfo}[2]{#2}
\providecommand{\BIBentrySTDinterwordspacing}{\spaceskip=0pt\relax}
\providecommand{\BIBentryALTinterwordstretchfactor}{4}
\providecommand{\BIBentryALTinterwordspacing}{\spaceskip=\fontdimen2\font plus
\BIBentryALTinterwordstretchfactor\fontdimen3\font minus
  \fontdimen4\font\relax}
\providecommand{\BIBforeignlanguage}[2]{{%
\expandafter\ifx\csname l@#1\endcsname\relax
\typeout{** WARNING: IEEEtran.bst: No hyphenation pattern has been}%
\typeout{** loaded for the language `#1'. Using the pattern for}%
\typeout{** the default language instead.}%
\else
\language=\csname l@#1\endcsname
\fi
#2}}
\providecommand{\BIBdecl}{\relax}
\BIBdecl

\bibitem{Arman3}
A.~Ahmadzadeh, H.~Arjmandi, A.~Burkovski, and R.~Schober, ``Reactive receiver
  modeling for diffusive molecular communication systems with molecule
  degradation,'' in \emph{Proc. IEEE ICC}, May 2016, pp. 1--7.

\bibitem{NakanoB}
T.~Nakano, A.~W. Eckford, and T.~Haraguchi, \emph{Molecular
  Communication}.\hskip 1em plus 0.5em minus 0.4em\relax Cambridge University
  Press, 2013.

\bibitem{AlbertsBook}
B.~Alberts, D.~Bray, K.~Hopkin, A.~D. Johnson, A.~Johnson, J.~Lewis, M.~Raff,
  K.~Roberts, and P.~Walter, \emph{Essential Cell Biology}, 3rd~ed.\hskip 1em
  plus 0.5em minus 0.4em\relax Garland Science, 2009.

\bibitem{PierobonJ1}
M.~Pierobon and I.~Akyildiz, ``Diffusion-based noise analysis for molecular
  communication in nanonetworks,'' \emph{{IEEE} Trans. Signal Process.},
  vol.~59, no.~6, pp. 2532--2547, Jun. 2011.

\bibitem{PierobonJ2}
------, ``A statistical-physical model of interference in diffusion-based
  molecular nanonetworks,'' \emph{{IEEE} Trans. Commun.}, vol.~62, no.~6, pp.
  2085--2095, Jun. 2014.

\bibitem{MahfuzJ1}
M.~Mahfuz, D.~Makrakis, and H.~Mouftah, ``A comprehensive study of
  sampling-based optimum signal detection in concentration-encoded molecular
  communication,'' \emph{{IEEE} Trans. Nanobiosci.}, vol.~13, no.~3, pp.
  208--222, Sep. 2014.

\bibitem{PierobonJ3}
M.~Pierobon and I.~Akyildiz, ``Noise analysis in ligand-binding reception for
  molecular communication in nanonetworks,'' \emph{{IEEE} Trans. Signal
  Process.}, vol.~59, no.~9, pp. 4168--4182, Sep. 2011.

\bibitem{ShahMohammadianProc1}
H.~ShahMohammadian, G.~Messier, and S.~Magierowski, ``Modelling the reception
  process in diffusion-based molecular communication channels,'' in \emph{Proc.
  IEEE ICC}, Jun. 2013, pp. 782--786.

\bibitem{YilmazL1}
H.~Yilmaz, A.~Heren, T.~Tugcu, and C.-B. Chae, ``Three-dimensional channel
  characteristics for molecular communications with an absorbing receiver,''
  \emph{{IEEE} Commun. Lett.}, vol.~18, no.~6, pp. 929--932, Jun. 2014.

\bibitem{Akkaya}
A.~Akkaya, H.~B. Yilmaz, C.~B. Chae, and T.~Tugcu, ``Effect of receptor density
  and size on signal reception in molecular communication via diffusion with an
  absorbing receiver,'' \emph{{IEEE} Commun. Lett.}, vol.~19, no.~2, pp.
  155--158, Feb. 2015.

\bibitem{Heren}
A.~C. Heren, H.~B. Yilmaz, C.-B. Chae, and T.~Tugcu, ``Effect of degradation in
  molecular communication: Impairment or enhancement?'' \emph{IEEE Trans. Mol.
  Biol. Multi-Scale Commun.}, vol.~1, no.~2, pp. 217--229, Jun. 2015.

\bibitem{ChunJ1}
C.~T. Chou, ``Impact of receiver reaction mechanisms on the performance of
  molecular communication networks,'' \emph{{IEEE} Trans. Nanotechnol.},
  vol.~14, no.~2, pp. 304--317, Mar. 2015.

\bibitem{Yansha1}
Y.~Deng, A.~Noel, M.~Elkashlan, A.~Nallanathan, and K.~C. Cheung, ``Modeling
  and simulation of molecular communication systems with a reversible
  adsorption receiver,'' \emph{IEEE Trans. Mol. Biol. Multi-Scale Commun.},
  vol.~1, no.~4, pp. 347--362, Dec. 2015.

\bibitem{Berg1}
H.~Berg and E.~Purcell, ``Physics of chemoreception,'' \emph{Biophysical
  Journal}, vol.~20, no.~2, pp. 193 -- 219, Nov. 1977.

\bibitem{Zwanzig1}
R.~Zwanzig, ``Diffusion-controlled ligand binding to spheres partially covered
  by receptors: an effective medium treatment,'' \emph{Proceedings of the
  National Academy of Sciences}, vol.~87, no.~15, pp. 5856--5857, Aug. 1990.

\bibitem{Zwanzig2}
R.~Zwanzig and A.~Szabo, ``Time dependent rate of diffusion-influenced ligand
  binding to receptors on cell surfaces,'' \emph{Biophysical Journal}, vol.~60,
  no.~3, pp. 671 -- 678, Sep. 1991.

\bibitem{Berezhkovskii1}
A.~M. Berezhkovskii, Y.~A. Makhnovskii, M.~I. Monine, V.~Y. Zitserman, and
  S.~Y. Shvartsman, ``Boundary homogenization for trapping by patchy
  surfaces,'' \emph{The Journal of Chemical Physics}, vol. 121, no.~22, pp.
  11\,390--11\,394, Nov. 2004.

\bibitem{Berezhkovskii2}
A.~M. Berezhkovskii, M.~I. Monine, C.~B. Muratov, and S.~Y. Shvartsman,
  ``Homogenization of boundary conditions for surfaces with regular arrays of
  traps,'' \emph{The Journal of Chemical Physics}, vol. 124, no.~3, pp.
  036\,103--1--036\,103--3, Jan. 2006.

\bibitem{Steven_Andrews}
S.~S. Andrews and D.~Bray, ``Stochastic simulation of chemical reactions with
  spatial resolution and single molecule detail,'' \emph{Physical Biology},
  vol.~1, no.~3, p. 137, Aug. 2004.

\bibitem{NoelJ1}
A.~Noel, K.~C. Cheung, and R.~Schober, ``Improving receiver performance of
  diffusive molecular communication with enzymes,'' \emph{{IEEE} Trans.
  Nanobiosci.}, vol.~13, no.~1, pp. 31--43, Mar. 2014.

\bibitem{ChouJ1}
C.~T. Chou, ``Noise properties of linear molecular communication networks,''
  \emph{Nano Commun. Net.}, vol.~4, no.~3, pp. 87 -- 97, 2013.

\bibitem{GillespieBook}
D.~T. Gillespie and E.~Seitaridou, \emph{{Simple Brownian diffusion: an
  introduction to the standard theoretical models}}.\hskip 1em plus 0.5em minus
  0.4em\relax Oxford University Press, 2013.

\bibitem{GrinrodB1}
P.~Grindrod, \emph{The theory and applications of reaction-diffusion equations:
  patterns and waves}.\hskip 1em plus 0.5em minus 0.4em\relax Clarendon Press,
  1996.

\bibitem{H.KimJ1}
H.~Kim and K.~J. Shin, ``Exact solution of the reversible diffusion-influenced
  reaction for an isolated pair in three dimensions,'' \emph{Phys. Rev. Lett.},
  vol.~82, pp. 1578--1581, Feb. 1999.

\bibitem{Ivar}
I.~Stakgold and M.~J. Holst, \emph{Green's Functions and Boundary Value
  Problems}.\hskip 1em plus 0.5em minus 0.4em\relax Wiley, 2011.

\bibitem{SchultenL1}
K.~Schulten and I.~Kosztin, \emph{Lectures in Theoretical Biophysics},
  Champaign, IL, USA: University of Illinois at Urbana-Champaign, 2000.

\bibitem{WoldeJ1}
M.~J. Morelli and P.~R. ten Wolde, ``Reaction {B}rownian dynamics and the
  effect of spatial fluctuations on the gain of a push-pull network,'' \emph{J.
  Chem. Phys.}, vol. 129, no.~5, pp. 054\,112--1 -- 054\,112--11, 2008.

\bibitem{abramowitz}
M.~Abramowitz and I.~Stegun, \emph{Handbook of Mathematical Functions},
  1st~ed.\hskip 1em plus 0.5em minus 0.4em\relax New York: Dover, 1964.

\end{thebibliography}
\vspace*{-0.75 cm}
\begin{IEEEbiography}[{\includegraphics[width=1in,height=1.25in,clip,keepaspectratio]{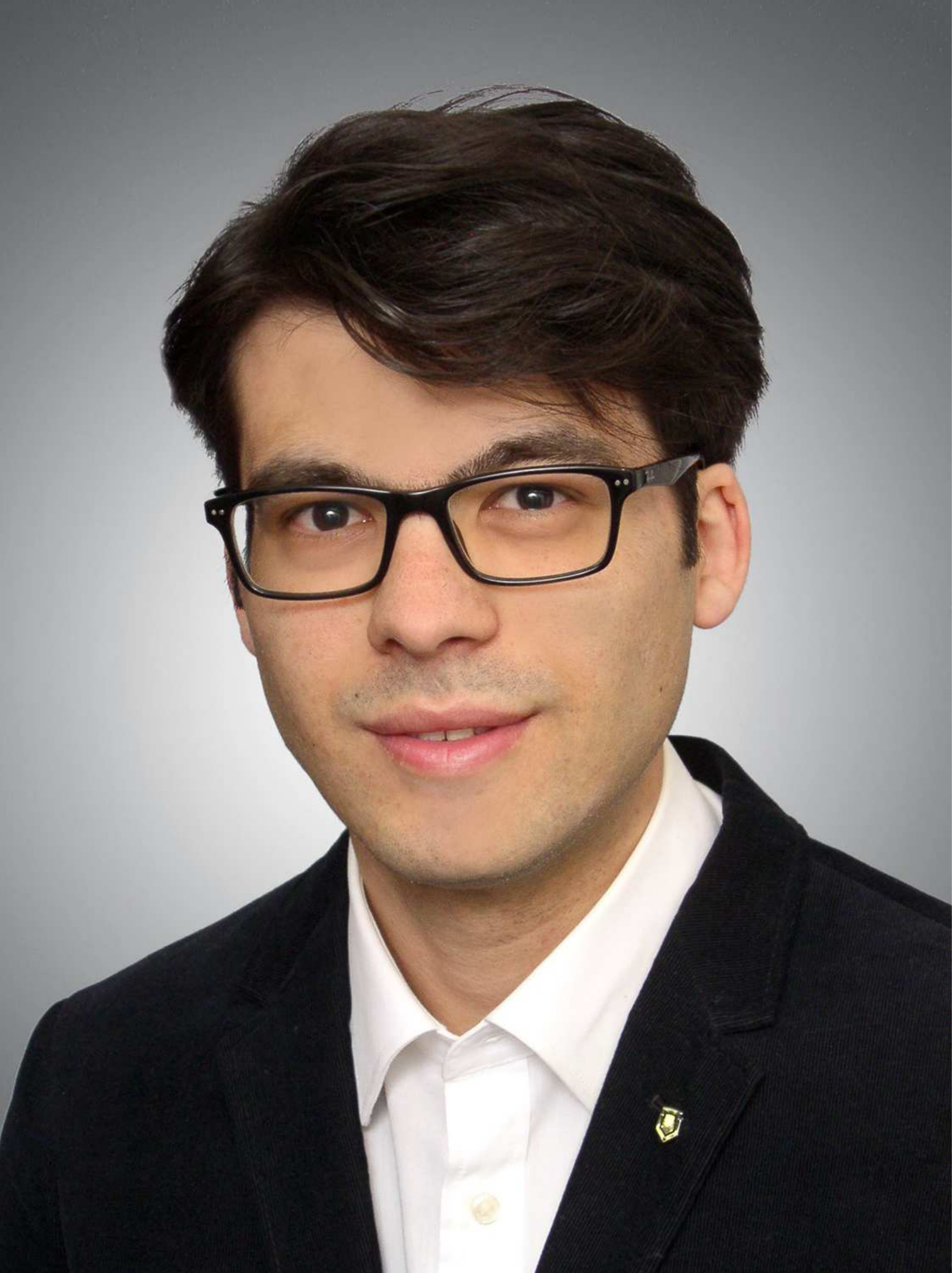}}]{Arman Ahmadzadeh} (S'14) received the B.Sc. degree in electrical engineering from the Ferdowsi University of Mashhad, Mashhad, Iran, in 2010, and the M.Sc. degree in communications and multimedia engineering from Friedrich-Alexander University, Erlangen, Germany, in 2013, where he is currently pursuing the Ph.D. degree in electrical engineering with the Institute for Digital Communications. His research interests include wireless communications and physical layer molecular communications. He received a best paper award from the IEEE International Conference on Communications (ICC) 2016.    
\end{IEEEbiography} 

\vspace*{-0.75 cm}
\begin{IEEEbiography}[{\includegraphics[width=1in,height=1.25in,clip,keepaspectratio]{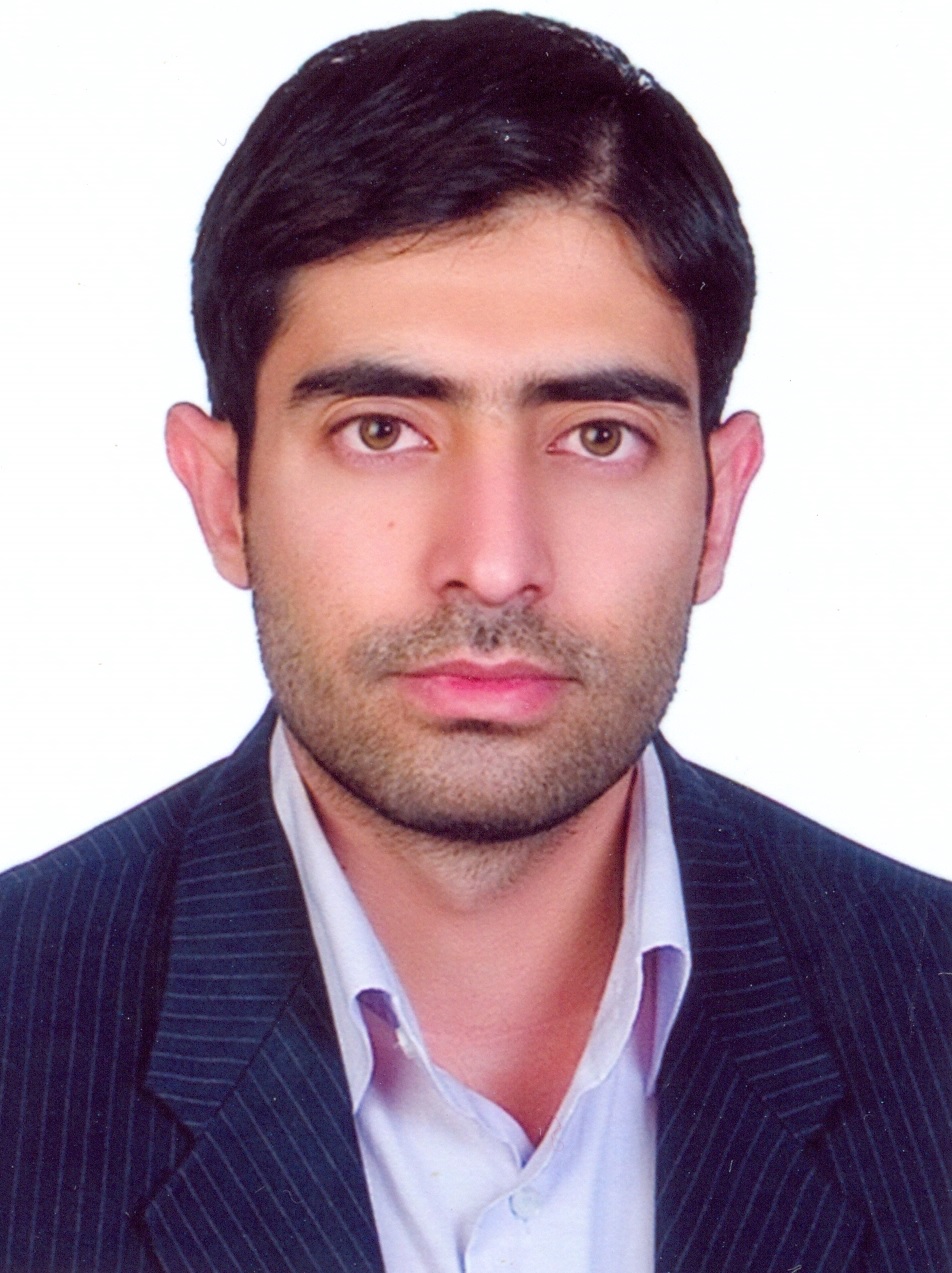}}]{Hamidreza Arjmandi} received the B.Sc. degree in electrical engineering from Shahed University, Tehran, Iran, in 2007, the M.Sc. degree in electrical engineering majoring in communication systems from the University of Tehran, Tehran, in 2010, and the PhD degree in electrical engineering from  Sharif University of Technology, Tehran, Iran. He is currently an assistant professor in the Department of Electrical Engineering at Yazd University, Yazd, Iran.  His main research interests are in nano-bio communications and wireless sensor networks.      
\end{IEEEbiography} 

\vspace*{-0.75 cm}
\begin{IEEEbiography}[{\includegraphics[width=1in,height=1.25in,clip,keepaspectratio]{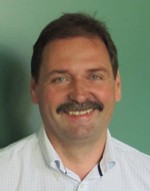}}]{Andreas Burkovski} (*1963) achieved his Diploma in Biology in 1989 and his Dr. rer. nat. degree in 1993, both at the University of Osnabrueck. After a postdoc period of one year at the University of Osnabrueck and four years at the Research Center Juelich (Institute of Biotechnology I), he became a group leader at the University of Cologne (Institute of Biochemistry) in 1997. In 2002 he completed his Habilitation at the University of Cologne and received the \textit{venia legendi} in Biochemistry. In 2005 he received a professorship for Microbiology at the University of Erlangen-Nuremberg. Andreas Burkovski was coordinator of a BMBF proteomics network and member of the scientific advisory board of the BMBF systems biology program SysMo2. He is member of the editorial board of BMC Microbiology and, since 2009, spokesman of the CRC796 integrated graduate school “Erlangen School of Molecular Communication”. 

Main research topics focus on the analysis of signalling networks in Gram-positive bacteria from single molecule to multi -omics level. In recent years, the focus of research on regulatory networks shifted towards synthetic biology and molecular communication aspects. As a further topic, host pathogen interactions are studied in his group. This project focuses on the identification of virulence factors of \textit{Corynebacterium diphtheriae} and other toxigenic corynebacteria, which are crucial for adhesion to host cells, invasion of epithelial cells and survival within epithelial cells and macrophages. Also in this project, signalling events are of major interest.      
\end{IEEEbiography}

\begin{IEEEbiography}[{\includegraphics[width=1in,height=1.25in,clip,keepaspectratio]{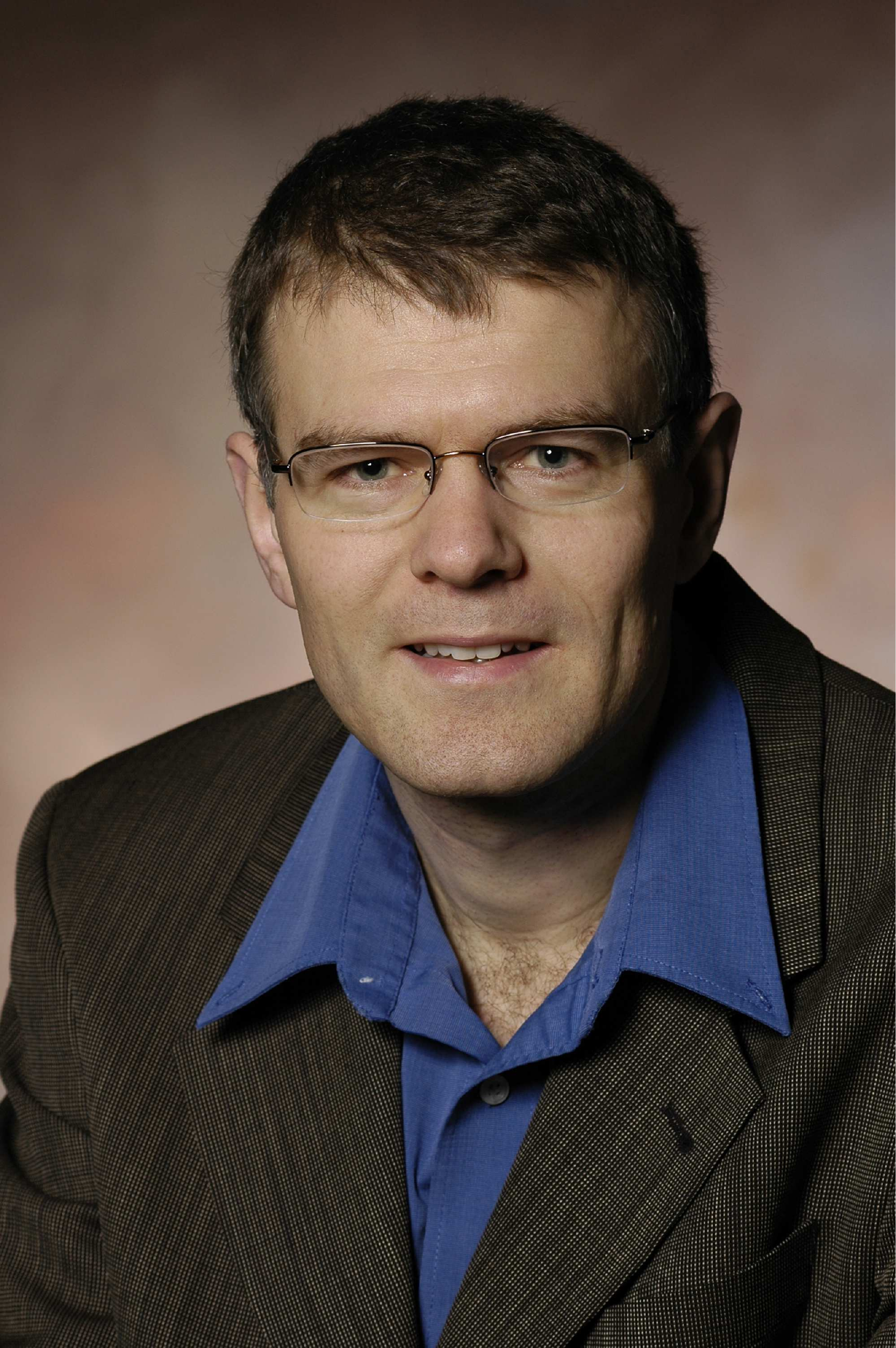}}]{Robert  Schober} (S'98, M'01, SM'08, F'10) was born in Neuendettelsau, 
Germany, in 1971. He received the Diplom (Univ.) and the Ph.D. degrees 
in electrical engineering from the University of Erlangen-Nuermberg in 
1997 and 2000, respectively. From May 2001 to April 2002 he was a 
Postdoctoral Fellow at the University of Toronto, Canada, sponsored by 
the German Academic Exchange Service (DAAD). Since May 2002 he has been 
with the University of British Columbia (UBC), Vancouver, Canada, where 
he is now a Full Professor. Since January 2012 he is an Alexander von 
Humboldt Professor and the Chair for Digital Communication at the 
Friedrich Alexander University (FAU), Erlangen, Germany. His research 
interests fall into the broad areas of Communication Theory, Wireless 
Communications, and Statistical Signal Processing.

Dr. Schober received several awards for his work including the 2002 
Heinz Maier Leibnitz Award of the German Science Foundation (DFG), the 
2004 Innovations Award of the Vodafone Foundation for Research in 
Mobile Communications, the 2006 UBC Killam Research Prize, the 2007 
Wilhelm Friedrich Bessel Research Award of the Alexander von Humboldt 
Foundation, the 2008 Charles McDowell Award for Excellence in Research 
from UBC, a
2011 Alexander von Humboldt Professorship, and a 2012 NSERC E.W.R. 
Steacie Fellowship. In addition, he received best paper awards from the 
German Information Technology Society (ITG), the European Association 
for Signal, Speech and Image Processing (EURASIP), the IEEE ICC 2016,  IEEE WCNC 2012, IEEE 
Globecom 2011, IEEE ICUWB 2006,  the International Zurich Seminar on 
Broadband Communications, and European Wireless 2000. Dr. Schober is a 
Fellow of the Canadian Academy of Engineering and a Fellow of the 
Engineering Institute of Canada. From 2012 to 2015 he served as 
Editor-in-Chief of the IEEE \textsc{Transactions on Communications}. He is 
currently the Chair of the Steering Committee of the IEEE \textsc{Transactions 
on Molecular, Biological and Multiscale Communication} and a 
Member-at-Large on the Board of Governors of the IEEE Communication Society.
\end{IEEEbiography}

\end{document}